\documentstyle[prb,aps,psfig]{revtex}

\begin{document}

\title{A microscopic model for d-wave charge carrier pairing \\ and
 non-Fermi-liquid behavior \\ in a purely repulsive 2D electron
 system}

\author{Mona Berciu and Sajeev John}

\address{ Department of Physics, University of Toronto, 60 St. George
Street, Toronto, Ontario, M5S 1A7, Canada}

\date{\today}

\maketitle


\begin{abstract}

We investigate a  microscopic model for  strongly correlated
electrons  with both on-site and nearest neighbor Coulomb repulsion 
 on a 2D square lattice. This exhibits a state
in which electrons undergo a ``somersault'' in their
internal spin-space (spin-flux) as they traverse a closed loop in
external coordinate space. 
 When this spin-1/2
antiferromagnetic (AFM) insulator is doped, the ground state
is a liquid of charged, bosonic meron-vortices, which for
topological reasons are created in  vortex-antivortex pairs.
The magnetic exchange energy of the distorted AFM background leads to
a logarithmic vortex-antivortex attraction which overcomes the direct
Coulomb repulsion between holes localized on the vortex cores. This
leads to the appearance of pre-formed charged pairs. 
 We use the Configuration Interaction (CI)
Method to study the quantum translational and rotational
motion of various charged magnetic solitons and soliton pairs.
 The CI method systematically
describes fluctuation and quantum tunneling corrections to the
Hartree-Fock Approximation (HFA).
We find that the lowest energy charged  meron-antimeron pairs exhibit
d-wave rotational symmetry, consistent with the symmetry of
the cuprate superconducting order parameter. For a single
hole in the 2D AFM plane, we find a precursor to spin-charge
separation in which a conventional charged spin-polaron dissociates
into a singly charged meron-antimeron pair. 
 This model provides a unified microscopic basis
for (i) non-Fermi-liquid transport properties, (ii) d-wave
preformed charged carrier pairs, (iii)  mid-infrared
optical absorption, (iv) destruction of AFM long range order with
doping and other magnetic properties, and (v) certain aspects of
angled resolved photo-emission spectroscopy (ARPES).

\end{abstract}

\narrowtext \twocolumn

\section{Introduction}

The microscopic understanding of the effect of charge carrier doping
on spin$-{1 \over 2}$ antiferromagnetic (AFM) Mott insulators is the
central issue of the high-temperature superconducting
cuprates. \cite{PWscience} Many puzzling experimental features of
these systems \cite{Ginz} suggest that a fundamental law of nature
remains to be recognized.  Extremely low doping ( $\delta \sim
0.02-0.05$ charge carriers per site) leads to a complete destruction
of the long-range AFM order, and a transition to an unusual
non-Fermi-liquid metal. This unusual metal becomes superconducting,
with the transition temperature $\mathrm{T_c}$ strongly dependent on
the doping $\delta$. The maximum $\mathrm{T_c}$ is reached for dopings
around $\delta=0.15$. For higher dopings the critical temperature
decreases to zero, and in the overdoped region a crossover towards a
(non-superconducting) Fermi-liquid takes place.  Two central questions
require resolution. The first one concerns the nature of the charge
carriers responsible for this non-Fermi-liquid metallic behavior. This
is a fundamental issue, since it lies outside the framework of
Landau's Fermi-liquid theory and it necessitates understanding the
appearance of non-quasiparticle-like charge carriers in a system of
interacting electrons.  The second question concerns the nature of
strong attractive pairing between these charge carriers, given the
purely repulsive interaction between the constituent electrons.  In
conventional superconductors, the pairing attraction is due to
overscreening of the electron-electron Coulomb repulsion by the ionic
lattice.  In the case of the high-temperature superconducting
cuprates, it has been suggested \cite{PWscience} that pairing is an
intrinsic property of the electron gas itself mediated by AFM
spin-fluctuations of the doped system. Accordingly, the challenge is
to identify a strong attractive force based purely on repulsive Coulomb
interactions. In this paper, we derive such a force and demonstrate
that it leads to d-wave pairing of charge carrying holes introduced by
doping a quantum, spin-${1 \over 2}$, Mott-Hubbard antiferromagnet.

The simplest model Hamiltonians used to investigate the cuprate
physics are the Hubbard model and the closely related t-J
model. Unlike the 1D problem, an exact solution for the 2D Hubbard
Hamiltonian is not known. As a result, various approximations are
necessary. Although the application of the mean-field theory has been
severely criticized in this context, it provides a valuable reference
point for incorporating fluctuation effects. Moreover, even for the 1D
Hubbard model, essential features of the exact solution may be
recaptured by judiciously incorporating fluctuation and tunneling
effects into mean-field theory. \cite{prev} The most straightforward
mean-field theory is the Hartree-Fock Approximation (HFA). At
half-filling ($\delta=0$) the HFA predicts an AFM Mott insulator
ground-state. As the system is doped, HFA suggests that charge carrier
holes in the AFM background assemble in charged stripes, which are
quasi-one-dimensional structures. \cite{16,16bb,16bis} A large effort
has been devoted to studying these charged stripes and relating them
to certain features of the cuprates. \cite{preformed}

Recently, a more fundamental investigation of the many-electron
problem has suggested the possibility of an alternative model
Hamiltonian for the cuprate physics. This model Hamiltonian, called
the spin-flux model, \cite{1} suggests that the long-range Coulomb
interaction between spin-${1 \over 2}$ electrons leads to qualitative
new physics, not apparent in the conventional Hubbard model (see
Section 2).  The results of the Hartree-Fock
study \cite{MB3} of this spin-flux model are summarized in
Section 3. They suggests that the undoped
parent compound is also an AFM Mott insulator. However, unlike the
conventional Hubbard model, the one-electron dispersion relations of
the AFM mean-field in the spin-flux model match those measured
experimentally through angle-resolved photo-emission for undoped
cuprates. A proper description of the highest occupied electronic
states (as provided by the spin-flux model), is crucial to
considerations of doping.  The spin-flux model and the conventional
Hubbard model differ dramatically in this regard.  At the HF level,
the doping holes added to the AFM background of the spin-flux model
are trapped in the core of antiferromagnetic spin vortices. This
composite object (meron-vortex) is a bosonic charged collective mode of the
many-electron system (the total spin of the magnetic vortex is
zero). The reversal of the spin-charge connection provides a
microscopic basis for non-Fermi-liquid behavior.

A magnetic vortex is strongly attracted to an antivortex. This
attraction increases logarithmically with the distance between the
vortex cores, and is stronger than the unscreened Coulomb repulsion
between the charge meron-vortex cores. In effect, the increase in
Coulomb energy between a given pair of holes is more than offset by
the lowering in exchange energy between the background electrons as their
vortices approach each other from far away. As the inter-vortex distance
increases, more and more spins are rotated out of their AFM background
orientation and the total energy of the system increases. Thus, even
at the HF level, the spin-flux model provides a fundamental
underpinning for the origin of both non-Fermi-liquid behavior, and
strong pairing between the charge carriers.

While providing a good starting point, the Hartree-Fock Approximation
also has serious shortcomings. For instance, the ground-state
wavefunction in the presence of doping is non-homogeneous (the static
meron-vortices of the spin-flux model, or the charged stripes of the
conventional Hubbard model, break translational symmetry).
Physically, one expects that these charge carriers can move along the
planes, resulting in a wavefunction which preserves the translational
symmetry of the original Hamiltonian. Quantum dynamics of the charge
carriers also determines whether the doped ground-state is really a
metal. Charge carriers in the optimally doped cuprates are quite
mobile excitations, although their scattering rates are radically
different from electrons in a conventional Fermi liquid.

A consistent way of treating the quantum dynamics of the charge
carriers is provided by the Configuration Interaction (CI)
Method,\cite{CIpaper,prev} described in Section 4. 
 Here, a linear combination of HF
wavefunctions is used in order to restore the various broken
symmetries. For instance, in a doped system the CI wavefunction is
chosen to be a linear combination of HF wavefunctions with the charge
carrier localized at different sites. Certain types of charge carriers
can lower their total energy substantially by quantum mechanically
hopping from one site to the next. We tested the accuracy of the CI
method against the exact 
solution \cite{wu} of the one-dimensional Hubbard model in Reference
3. In the 1D Hubbard model the CI
method describes the quantum dynamics of charged domain wall solitons
in the AFM background. By including these effects as
fluctuation corrections to the Hartree-Fock mean-field theory, the CI
method provides excellent agreement with the exact Bethe Ansatz
solution  for the ground-state energy of the doped 1D Hubbard
chain, over the entire $U/t$ range.  The CI
method also leads  to a clear
demonstration of the spin-charge separation in 1D. Addition of one doping
hole to the half-filled antiferromagnetic chain results in the
appearance of two different carriers: a charged bosonic domain-wall
(which carries the charge but no spin) and a neutral spin-1/2 domain
wall (which carries the spin but no charge). This study \cite{prev} 
demonstrates the
effectiveness of the CI method. In this paper we use the CI method to
investigate dynamics of the charged meron-vortices in the spin-flux
model. Throughout this paper we exploit and refer to the analogy
between the charge excitations of the 1D Hubbard model and the 2D
spin-flux model, \cite{MB1,MB2} apparent in the CI
approach. \cite{prev} The CI results for the spin-flux model
(presented in Section 4) confirm that the meron-vortices are very
mobile, suggesting that a collection of such mobile bosonic charge
carriers is a non-Fermi-liquid metal. The CI method also allows us to
identify the rotational symmetry of the meron-antimeron pair
wavefunction to be d-wave for the most stable pairs. An energetically
more expensive metastable s-wave pairing is also possible. The
possibility of spin-charge separation in 2D is elucidated.  A summary
of the results, their interpretation and conclusions is provided in
Section 5.

\section{The spin-flux model}

The effective 2D Hamiltonian that we use to describe the strongly
correlated electrons residing in the $O(2p)-Cu(3d_{x^2-y^2})$ orbitals
of the isolated CuO$_2$ plane is the tight-binding model:
\begin{equation}
\label{1.101}
{\cal H}=- \sum_{i,j, \sigma}^{} \left( t_{ij} c^{\dagger}_{i\sigma}
c_{j\sigma} + h.c.\right) + \sum_{i,j}^{} V_{ij} n_{i} n_{j}
\end{equation}
where $c_{i\sigma}^{\dagger}$ creates an electron at site $i$ with
spin $\sigma$, $t_{ij}$ is the hopping amplitude from site $j$ to site
$i$ on the square lattice, $n_i\equiv
\sum\limits^2_{\sigma=1}c_{i\sigma}^{\dagger}c_{i\sigma}$ is the total
number of electrons at site $i$, and $V_{ij}$ is the Coulomb
interaction between electrons at sites $i$ and $j$.  The dominant
terms are the nearest-neighbor hopping $t_{ij}=t_0$ and the on-site
Coulomb repulsion $V_{ii}=U/2$. If only these two terms are
considered, and we shift the chemical potential by $U$, this reduces
to the well-known Hubbard model.  The neglect of all longer range
Coulomb interaction ($V_{ij}=0$, if $i\neq j$), in the Hubbard model,
is based on the Fermi-liquid theory notion of screening of the
effective electron-electron interaction.  However, Fermi liquid theory
fails to explain many of the crucial features of the high-T$_c$
cuprates. In our description, we include the nearest-neighbor Coulomb
repulsion, which we assume is on the energy scale of $t$. This leads
to the generation of spin-flux, an entirely new type of broken
symmetry in the many-electron system, which we show leads naturally to
bosonic charge carriers in the form of meron-vortices,
non-Fermi-liquid behavior and a strong attractive pairing force
between holes in the AFM background. In order to extract the relevant
physics from our starting Hamiltonian,
\begin{equation}
\label{1.103}
{\cal H}\!=\!-t_0\!\! \sum_{\!\!\langle i,j\rangle \sigma}^{} \left(
c^{\dagger}_{i\sigma} c_{j\sigma} \!+\! h.c.\!\right)\! +\! U
\!\sum_{i}^{} \!n_{i\uparrow} n_{i\downarrow}\!+\! V \!\!\sum_{\langle
i,j\rangle}^{} n_{i} n_{j}
\end{equation}
we introduce bilinear combination of electron operators $
\Lambda^\mu_{ij}\equiv
c^{\dagger}_{i\alpha}\sigma^\mu_{\alpha\beta}c_{j\beta}$, $\mu=0, 1,
2, 3$, for $i\neq j$ (summation over multiple indexes is assumed).
Here $\sigma^0$ is the $2\times 2$ identity matrix and $\vec
\sigma\equiv (\sigma^1, \sigma^2,\sigma^3)$ are the usual Pauli spin
matrices. The notation $\langle i, j \rangle$ means that the sites $i$
and $j$ are nearest neighbors.  The quantum expectation value
$\langle\ \rangle$ of the $\Lambda^{\mu}_{ij}$ operators are
associated with charge-currents ($\mu=0$) and spin-currents
($\mu=1,2,3$).  Non-vanishing charge currents lead to appearance of
electromagnetic fields, which break the time-reversal symmetry of the
Hamiltonian. Experimentally, this does not occur in the cuprates.  In
the following, we adopt the ansatz that there is no charge current in
the ground state $\Lambda^{0}_{ij}= 0$ but circulating spin-currents
may arise and take the form $\Lambda^{a}_{ij} = {2t_0\over V}
i\Delta_{ij}\hat n_a, a= 1,2,3$, where $\vert \Delta_{ij}\vert
=\Delta$ for all $i$ and $j$, and $\hat{n}$ is a unit vector. These
spin-currents provide a transition state to the spin-flux mean-field
that we use in this paper.

\begin{figure}[h]
\centering
\parbox{0.45\textwidth}{\psfig{figure=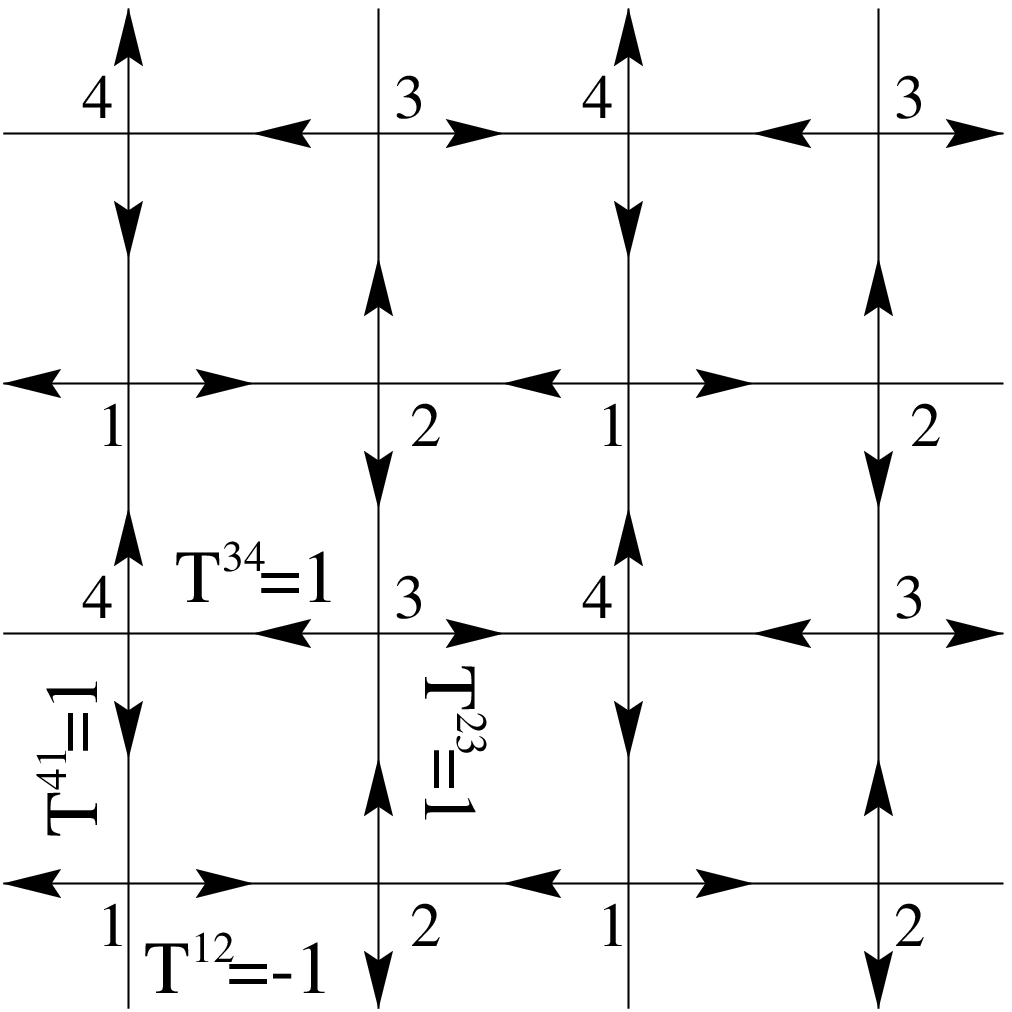,width=55mm}}
\caption{\label{fig1.55} Choice of the gauge for describing the
mean-field spin-flux background. Physical observables depend on the
rotation matrices $T^{ij}$ only through the plaquette matrix product
$T^{12}T^{23}T^{34}T^{41}$. Shown above is the simplest (spin
independent) gauge choice describing a $2\pi$-rotation of the internal
coordinate system of the electron (described by 3 Euler angles) as it
encircles an elementary plaquette. This is a new form of spontaneous
symmetry breaking for a strongly interacting electron system in which
the mean-field Hamiltonian acquires a term with the symmetry of a
spin-orbit interaction.  }
\end{figure}

Using the Pauli spin-matrix identity, ${1\over2}
\sigma^\mu_{\alpha\beta}(\sigma^\mu_{\alpha^\prime\beta^\prime})^ \ast
=\delta_{\alpha\alpha^\prime} \delta_{\beta\beta^\prime}$, it is
possible to rewrite the nearest-neighbor electron-electron interaction
terms as $n_in_j=2n_i -{1 \over 2}
\Lambda^\mu_{ij}(\Lambda_{ij}^\mu)^+$. If we neglect fluctuations in
the spin-currents, we can use the mean-field factorization,
$\Lambda^\mu_{ij}(\Lambda^\mu_{ij})^+\rightarrow \langle
\Lambda^\mu_{ij}\rangle (\Lambda^\mu_{ij})^+ +\Lambda^\mu_{ij} \langle
\Lambda^\mu_{ij}\rangle^* -\langle\Lambda^\mu_{ij}\rangle \langle
\Lambda^\mu_{ij}\rangle^*$.  Thus, the quartic nearest-neighbor
Coulomb interaction term is reduced to a quadratic term that is added
to the hopping term leading to the effective Hamiltonian:
\begin{equation}
\label{1.104}
{\cal H}=-t\sum_{\langle i,j \rangle \atop \alpha\beta}^{} \left (
c^{\dagger}_{i\alpha} T^{ij}_{\alpha\beta} c_{j\beta} + h.c. \right) +
U \sum_{i}^{} n_{i\uparrow} n_{i\downarrow}.
\end{equation}
 Here, $T^{ij}_{\alpha\beta}\equiv (\delta_{\alpha\beta}
+i\Delta_{ij}\hat{n}\cdot\vec \sigma_{\alpha\beta})/\sqrt{1+\Delta^2}$
are spin-dependent $SU(2)$ hopping matrix elements defined by the
mean-field theory, and $t=t_o\sqrt{1+\Delta^2}$.  In deriving
(\ref{1.104}) we have dropped constant terms which simply change the
zero of energy as well as terms proportional to $\sum\limits_in_i$
which simply change the chemical potential. It was shown previously
\cite{1,2} that the ground state energy of the Hamiltonian of
Eq.(\ref{1.104}) depends on the SU(2) matrices $T^{ij}$ only through
the plaquette matrix product $T^{12}T^{23}T^{34}T^{41}\equiv\exp
(i\hat n\cdot\vec\sigma\Phi)$.  Here, $\Phi$ is the spin-flux which
passes through each plaquette and $2\Phi$ is the angle through which
the internal coordinate system of the electron rotates as it encircles
the plaquette.  Since the electron spinor wavefunction is two-valued,
there are only two possible choices for $\Phi$.  If $\Phi=0$ we can
set $T^{ij}_{\alpha\beta}=\delta_{ij}$ and the Hamiltonian
(\ref{1.104}) describes conventional ordered magnetic states of the
Hubbard model.  The other possibility is that a spin-flux $\Phi=\pi$
penetrates each plaquette, leading to
$T^{12}T^{23}T^{34}T^{41}=-1$. This means that the one-electron
wavefunctions are antisymmetric around each of the plaquettes,
i.e. that as an electron encircles a plaquette, its wavefunction in
the internal spin space of Euler angles rotates by $2\pi$ in response
to strong interactions with the other electrons.  In effect, the
electron performs an internal ``somersault'' as it traverses a closed
path in the CuO$_2$ plane.  This spin-flux phase is accompanied by a
AFM local moment background (with reduced magnitude relative to the
AFM phase of the conventional Hubbard model). In the spin-flux phase,
the kinetic energy term in (\ref{1.104}) exhibits broken symmetry as
though a spin-orbit interaction has been added. However, it is
distinct from the smaller, conventional spin-orbit effects which give
rise to anisotropic corrections to superexchange interactions between
localized spins in the AFM. \cite{7bis} In the presence of charge
carriers this mean-field is unstable to the proliferation of
topological fluctuations (magnetic solitons) which eventually destroy
AFM long range order. In this sense, the analysis which we present
below goes beyond simple mean field theory. The quantum dynamics of
these magnetic solitons described by the Configuration Interaction (CI)
method, corresponds to tunneling effects not contained in the
Hartree-Fock approximation.  For simplicity, throughout this paper we
assume that the mean-field spin-flux parameters $T^{ij}$ are given by
the simplest possible choice $T^{12}=-1, T^{23}=T^{34}=T^{41}=1$ (see
Fig. \ref{fig1.55}). In order to go beyond a mean-field description of
the spin-flux, these matrices may also be treated as dynamical
variables. In this paper, we go beyond mean-field theory in describing
the antiferromagnetic degrees of freedom but restrict ourselves to a
mean-field model of the spin-flux.

\section{Hartree-Fock results for the spin-flux model}

The Configuration Interaction Method utilizes a linear combination of
judiciously chosen Hartree-Fock wavefunctions.\cite{CIpaper,prev} In
this section, we provide a short review of the relevant Hartree-Fock
results for the spin-flux model. A full comparison between the HFA for
the spin-flux model and the conventional Hubbard model has been
published elsewhere. \cite{MB3}

\subsection{The Static Hartree-Fock Approximation}

One of the most widely used approximations for the many-electron
problem is the Static Hartree-Fock Approximation (HFA).  In this
approximation the many-body problem is reduced to one-electron
problems in which each electron moves in a self-consistent manner
depending on the mean-field potential of the other electrons in the
system.  While this method is insufficient, by itself, to capture all
of the physics of low dimensional electronic systems with strong
correlations, it provides a valuable starting point from which
essential fluctuation corrections can be included. In particular, we
use the Hartree-Fock method to establish the electronic structure and
the static energies of various magnetic soliton structures. In the
more general Configuration Interaction (CI) variational wavefunction,
the solitons acquire quantum dynamics and describe large amplitude
tunneling and fluctuation effects that go beyond mean field theory.

In the HF approximation, the many-body wavefunction $| \Psi\rangle$ is
decomposed into a Slater determinant of effective one-electron
orbitals. The one-electron orbitals are found from the condition that
the total energy of the system is minimized
\begin{equation}
\label{2.1}
\delta { \langle \Psi| {\cal H}| \Psi\rangle \over \langle\Psi |
\Psi\rangle} =0
\end{equation}

In order to approximate the ground state of the spin-flux 
Hamiltonian (\ref{1.104}), we
consider a Slater determinant trial-wavefunction of the form
\begin{equation}
\label{2.3}
|\Psi\rangle=\prod_{p=1}^{N_e} a^{\dagger}_p |0\rangle,
\end{equation}
where $|0\rangle$ is the vacuum state, $N_e$ is the total number of
electrons in the system and the one-electron states are given by
\begin{equation}
\label{2.4}
a^{\dagger}_n=\sum_{i \sigma}^{} \phi_n(i,\sigma)
c^{\dagger}_{i\sigma}
\end{equation}
Here, the one-particle wave-functions $\phi_n(i,\sigma)$ form a
complete and orthonormal system.

Using the wavefunction (\ref{2.3}) in equation (\ref{2.1}), and
minimizing with respect to the one-particle wavefunctions
$\phi_n(i,\sigma)$, we obtain the Hartree-Fock eigen-equations:
\parbox{80mm}{
$$ E_n \phi_n(i,\alpha)= -t \sum_{j \in V_i, \beta}
T^{ij}_{\alpha\beta} \phi_n(j,\beta)$$
\begin{equation} 
\label{2.5}
+U \sum_{\beta}^{} \left( {1 \over 2}\delta_{\alpha\beta} Q(i)
-\vec{\sigma}_{\alpha\beta} \vec{S}(i)\right) \phi_n(i,\beta)
\end{equation}}
where $(\sigma_x, \sigma_y, \sigma_z)$ are the Pauli spin matrices and
the charge density,
\begin{equation}
\label{2.6}
Q(i)=\langle\Psi|c^{\dagger}_{i\alpha}c_{i\alpha}|\Psi\rangle=
\sum_{p=1}^{N_e} |\phi_p(i,\alpha)|^2,
\end{equation}
and the spin density,
\begin{equation}
\label{2.7}
\vec{S}(i)=\langle\Psi|c^{\dagger}_{i\alpha} {
\vec{\sigma}_{\alpha\beta} \over 2} c_{i\beta}|\Psi\rangle = \sum_{p=1
}^{N_e} \phi^*_p(i,\alpha){ \vec{\sigma}_{\alpha\beta} \over 2}
\phi_p(i,\beta),
\end{equation}
 must be computed self-consistently.  The notation $j \in V_i$
appearing in (\ref{2.5}) means that the sum is performed over the
sites $j$ which are nearest-neighbors of the site $i$. The
self-consistent Hartree-Fock equations (\ref{2.5},\ref{2.6},\ref{2.7})
must be satisfied by the occupied orbitals $p=1\dots N_e$, but can
also be used to compute the empty (hole) orbitals.

The ground-state energy of the system in the HFA is given by
\begin{equation}
\label{2.8}
E_{GS}=\langle \Psi| {\cal H} |\Psi\rangle= \sum_{p=1}^{N_e} E_p - U
\sum_{i}^{}\left({1\over 4} Q(i)^2- \vec{S}(i)^2\right)
\end{equation}
where the single particle energies are obtained from (\ref{2.5}).

The approximation scheme described above is called the Unrestricted
Hartree-Fock Approximation, because we did not impose constraints on
the wavefunction $| \Psi\rangle$ which would require it to be an
eigenfunction of various symmetry operations which commute with the
Hamiltonian (\ref{1.104}). If these symmetries are enforced, the method
is called the Restricted Hartree-Fock Approximation.  We use the
Unrestricted HFA since it leads to lower energies. The breaking of
symmetries in our case implies that electronic correlations are more
effectively taken into account. \cite{Fulde} The restoration of these
symmetries is deferred until the CI wavefunction is introduced.
 
In the undoped (half-filled) case, the self-consistent Hartree-Fock
equations can be solved analytically for the infinite system, using
plane-wave one-particle wave-functions.  In the unrestricted
Hartree-Fock approach, doping the system leads to the appearance of
inhomogeneous solutions, which break the translational invariance.  In
this case, we solve the unrestricted self-consistent Hartree-Fock
equations numerically on a finite lattice.  Starting with an initial
spin and charge distribution $\vec{S}(i)$ and $Q(i)$, we numerically
solve the eigenproblem (\ref{2.5}) and find the HF eigenenergies $E_n$
and wavefunctions $\phi_n(i,\alpha)$. These are used in
Eqs. (\ref{2.6}) and (\ref{2.7}) to calculate the new spin and charge
distribution, and the procedure is repeated until self-consistency is
reached.  Numerically, we define self-consistency by the condition
that the largest variation of any of the charge or spin components on
any of the sites of the lattice is less that $10^{-9}$ between
successive iterations.

\subsection{The undoped ground state}

For the undoped system, the Hartree-Fock equations  (\ref{2.5}) for
an infinite system are easily solved. In the cuprates, long-range AFM
order is experimentally observed. Accordingly, we choose a spin
distribution at the site $i=\vec{e}_x i_x a+ \vec{e}_y i_y a$ of the
form $\vec{S}(i)= (-1)^{(i_x+i_y)}S \vec{e}$, where $\vec{e}$ is the
unit vector of some arbitrary direction, while the charge distribution
is $Q(i)=1$.  In the spin-flux phase, it is convenient to choose a
square unit cell, in order to simplify the description of the $T^{ij}$
phase-factors. We make the simplest gauge choice compatible with the
spin-flux condition for the $T$-matrices, namely that
$-T^{12}=T^{23}=T^{34}=T^{41}=1$ (see Fig. (\ref{fig1.55})).  This
leads to the reduced square Brillouin zone $-\pi/2a \le k_x \le
\pi/2a, -\pi/2a \le k_y \le \pi/2a$.  From the Hartree-Fock equations
we find two electronic bands, characterized by the dispersion
relations:
\begin{equation}
\label{2.27}
E^{(\pm)}_{sf}(\vec{k})=\pm E_{sf}(\vec{k})= \pm
\sqrt{\epsilon^2_{sf}(\vec{k})+(US)^2}
\end{equation}
where each level is four-fold degenerate and $\epsilon_{sf}(\vec{k})
=-2t \sqrt{\left(\cos{(k_xa)}\right)^2+\left( \cos{(k_ya)}\right)^2}$
are the noninteracting electron dispersion relations in the presence
of spin-flux.  The HF ground-state energy of the spin-flux AFM
background is given by (see Eq. (\ref{2.8})):
\begin{equation}
\label{2.28}
E_{GS}^{sf}=-4\sum_{\vec{k}} E_{sf}(\vec{k})+N^2 U\left(S^2+{1\over 4}
\right)
\end{equation}
where the AFM local moment amplitude is determined by the
self-consistency condition (\ref{2.7}) 
\begin{equation}
\label{2.29}
S={2\over N^2}\sum_{\vec{k}}^{} {US \over E_{sf}(\vec{k})}.
\end{equation}

\begin{figure}
\centering
\parbox{0.45\textwidth}{\psfig{figure=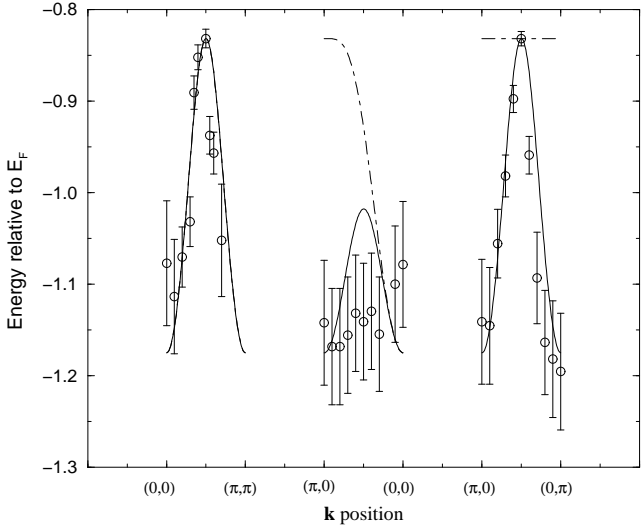,width=85mm}}
\caption{\label{fig2.13} A comparison between the experimentally
determined $E(\vec{k})$ quasi-particle dispersion relation, from angle
resolved photoemission studies (ARPES), for the insulating
Sr$_2$CuO$_2$Cl$_2$ (open circles with error bars) and the HF AFM
spin-flux model dispersion relation (full line) and the HF AFM
conventional Hubbard model dispersion relation (dashed-dotted line).
Three directions in $\vec{k}-$space are shown: $(0,0)$ to $(\pi,\pi)$,
$(\pi,0)$ to $(0,0)$ and $(\pi,0)$ to $(0,\pi)$.  While the peak on
the $(0,0)$ to $(\pi,\pi)$ is equally well described in both models,
the mean-field spin-flux model gives a much better agreement for the
$(\pi,0)$ to $(0,0)$ and $(\pi,0)$ to $(0,\pi)$ directions.  The
fitting corresponds to $U=2.01$ eV, $t=0.29$ eV for the spin-flux
phase, and $U=1.98$ eV, $t=0.21$ eV in the conventional phase. The
experimental results are the ARPES results of Ref. 17. }
\end{figure}

	At half-filling the valence band ($E_{sf}^{(-)}(k) <0$) is
completely filled, the conduction band ($E_{sf}^{(+)}(k) > 0$) is
completely empty, and a Mott-Hubbard gap of magnitude $2US$ opens
between the valence and the conduction bands. The ground-state of the
undoped spin-flux model is an AFM Mott insulator.  It is interesting
to note that the quasi-particle dispersion relation obtained in the
presence of the spin-flux (Eq. (\ref{2.27})) closely resembles the
dispersion as measured through angle-resolved photo-emission
spectroscopy (ARPES) in a compound such as Sr$_2$CuO$_2$Cl$_2$
\cite{Wells} (see Fig. \ref{fig2.13}).  There is a a large peak
centered at $(\pi/2, \pi/2)$ with an isotropic dispersion relation
around it, observed on both the $(0,0)$ to $(\pi,\pi)$ and $(0,\pi)$
to $(\pi,0)$ lines. The spin-flux model in HFA exhibits another
smaller peak at $(0,\pi/2)$ which is not resolvable in existing
experimental data. This minor discrepancy may be due to next nearest
neighbor hopping or other aspects of the electron-electron interaction
which we have not yet been included in our model. \cite{11} The
quasi-particle dispersion relation of the conventional Hubbard model
($T^{12}=T^{23}=T^{34}=T^{41}=1$) has a large peak at $(\pi/2,\pi/2)$
on the $(0,0)$ to $(\pi,\pi)$ line (see Fig. \ref{fig2.13}), but it is
perfectly flat on the $(0,\pi)$ to $(\pi,0)$ line (which is part of
the large nested Fermi surface of the conventional 2D Hubbard model).
Also, it has a large crossing from the upper to the lower band-edge on
the $(0,0)$ to $(0,\pi)$ line. This dispersion relation is very
similar to that of the $t-J$ model (see Ref. 17).

While both the conventional and spin-flux model predict AFM insulators
at half-filling (at least at the HF level), the spin-flux also
provides a much better agreement with the dispersion relations, as
measured by ARPES. As in the 1D case, \cite{prev} the effect of doping
is the appearance of discrete levels deep inside the Mott-Hubbard
gap. These levels are drawn into the gap from the top (bottom) of the
undoped valence (conduction) bands. Accordingly, the type of
excitations created by doping depends strongly on topology of the electronic
structure near the band edges.

\subsection{Charged solitons in the doped insulator: the spin-bag and
the meron-vortex}

\subsubsection{The spin-bag}

If we introduce just one hole in the plane, the self-consistent HFA
solution is a conventional spin-polaron or ``spin-bag'' (see Fig. 3a).
This type of excitation is the 2D analog of the 1D
spin-polaron. \cite{prev}  The
doping hole is localized around a particular site, leading to the
appearance of a small ferromagnetic core around that site. The spin
and charge distribution at the other sites are only slightly
affected. In fact, the localization length of the charge depends on
$U/t$, and becomes very large as $US \rightarrow 0$, since in this
limit the Mott-Hubbard gap closes.  For intermediate and large $U/t$,
the doping hole is almost completely localized on the five sites of
the ferromagnetic core.

	The spin-bag is a charged fermion, as can be seen by direct
inspection of its charge and spin distributions.  This is also
confirmed by its electronic structure (see Fig.3b). Thus, the 2D
spin-bag is indeed the analog of the 1D spin-polaron. \cite{prev}

\subsubsection{The meron-vortex}

The 2D analog of the 1D charged domain wall is the meron-vortex (see
Fig. \ref{fig2.22}).  Like the 1D domain-wall, the meron-vortex is
also a topological excitation, characterized by a topological
(winding) number $\pm 1$ (the spins on each sublattice rotate by $\pm
2\pi$ on any closed contour surrounding the center of the meron). As
such, a single meron-vortex cannot be created in an extended AFM
background with cyclic boundary conditions by the introduction of a
single hole (just as a single isolated charged domain wall cannot be
created on an AFM (even) chain with cyclic boundary conditions, by the
introduction of a single hole).  From a topological point of view,
this is so because the AFM background has a winding number 0, and the
winding number must be conserved, unless topological excitations
migrate over the boundary into the considered region.  However, 
excitations can be created in pairs of total topological
number 0. In the 1D case, this means creation of pairs of domain
walls, while in 2D this means the creation of vortex-antivortex pairs.

	From Figs. 4(a) and 4(b) we can see that the meron-vortex is a
charged boson. The total spin of such a configuration is zero, while
it carries the doping charge trapped in the vortex core.
  Moreover, from its electronic spectrum
(Fig. 4b) we can see that only the extended states of the valence band
are occupied. They are the only ones contributing to the total
spin. Since only one state is drawn from the valence band into the
gap, to become a discrete bound level, it appears that an odd
(unpaired) number of states remains in the valence band. However, one
must remember that for topological reasons, merons must appear in
vortex-antivortex pairs. Therefore, the valence band has an even
number of (paired) levels, and the total spin is zero. This argument
for the bosonic character of the meron-vortex is identical to that for
the charged domain wall in polyacetylene. \cite {MB2,polrev,SSH}

Unlike in the 1D case,\cite{prev} we cannot directly compare the
excitation energy of the spin-bag with the excitation energy of the
meron-vortex. The reason is that the excitation energy of the latter 
increases logarithmically with the size of the sample,
and therefore an isolated meron-vortex is always energetically more
expensive than a spin-bag. However, topology requires that merons and
antimerons are created in pairs. The excitation energy of such a
meron-antimeron pair is finite, allowing a meaningful comparison
between excitation energies of a pair of spin-bags and a
meron-antimeron pair.

\subsubsection{The meron-antimeron pair}

 In Figs.  5(a) and 5(b) we show the self-consistent spin and charge
 distributions for the lowest energy self-consistent HF configuration
 found when we add 2 holes to the AFM background, in the spin-flux
 model, for $U/t=5$.  This configuration consists of a meron and an
 antimeron centered on second nearest-neighbor sites. As a result of
 interactions, the cores of the vortices are somewhat distorted. If
 the vortices were uncharged, vortex-antivortex pair annihilation
 would be possible.  However, for charged vortices, the fermionic
 nature of the underlying electrons prevents two holes from being
 localized at the same site, in spite of the bosonic character of the
 collective excitation.

A very interesting feature of this configuration is the strong
topological attraction between the vortex and the antivortex.  The
closer the two cores are to each other, the smaller is the region in
which the spins are rotated out of their background AFM orientation by
the vortices, and therefore the smaller is the excitation energy of
the pair.  Since the holes are localized in the cores of the vortices,
this topological attraction between vortices is an effective
attraction between holes in the purely repulsive 2D electron
system. This effect is unique to the spin-flux phase. In the
conventional Hubbard model, vortices are not stable excitations. The
vortex-antivortex attraction increases as the logarithm of the
distance between the cores.  Therefore, the pair of vortices should
remain bound even if full unscreened $1/r$ Coulomb repulsion exists
between the charged cores,  providing a compelling scenario for the
existence of strongly bound pre-formed pairs in the underdoped regime.

There is another possible self-consistent state for the system with
two holes, consisting of two spin-bags far from each other (such that
their localized wave functions do not overlap). The excitation energy
of such a pair of spin-bags is simply twice the excitation energy of a
single spin-bag.  When this excitation energy is compared to the
excitation energy of the tightly bound meron-antimeron pair, we find
that it is higher by $0.15t$ (for $U/t=5$). In fact, for $U/t < 8$ the
HFA predicts that the meron-antimeron pair is the low-energy charged
excitation, while for $U/t >8$, the spin-bag is the low-energy charge
carrier. This is analogous with the situation in 1D, where the
spin-bag was predicted to be the low energy excitation for $U/t>6.5$,
in the HFA. \cite{prev} As in 1D, however, we expect that this
conclusion will be drastically modified once the charged solitons are
allowed to move along the planes and the lowering of kinetic energy
through translations is also taken into consideration.

We complete this review of the HF results by pointing out that the
strong analogy between the 1D Hubbard model and the 2D spin-flux model
is due to the similarity between the electronic structures at zero
doping. As seen from Fig.\ref{fig2.13}, the 2D spin-flux model has
isotropic dispersion relations about the $(\pi/2,\pi/2)$ point. This
acts as a Fermi point for the noninteracting system as it does in the
1D system.  The two empty discrete levels drawn deep inside the
Mott-Hubbard gap in the presence of the meron-vortex split from the
$(\pi/2,\pi/2)$ peaks of the electron dispersion relation. The
different topology of the large nested Fermi surface of the
conventional Hubbard model leads to instability of the meron-antimeron
pair. In fact, in the conventional Hubbard model doping holes assemble
in charged stripes, as opposed to the liquid of meron-antimeron pairs
which is the low-energy state of the doped spin-flux model.

\section{Configuration Interaction Method results for the 2D system}

\subsection{Configuration Interaction  Method}

  The essence of the Configuration Interaction (CI) method is that
the ground-state wavefunction, for a system with $N_e$ electrons, is
not represented by 
just a single $N_e\times N_e$ Slater determinant (as in the HFA), but a
judiciously chosen linear combination of such Slater
determinants. \cite{CIpaper} Given the fact that the set of all
possible Slater determinants (with all possible occupation numbers)
generated from a complete set of one-electron orbitals constitute a
complete basis of the $N_e$-particle Hilbert space, our aim is to pick
out a subset of Slater determinants which captures the essential
physics of the exact solution.

Consider the CI ground-state wavefunction given by
\begin{equation}
\label{3.1}
|\Psi\rangle= \sum_{i=1}^{N}\alpha_i|\Psi_i\rangle
\end{equation}
where each $|\Psi_i\rangle$ is a distinct $N_e\times N_e$ Slater
determinant and the coefficients $\alpha_i$ are chosen to satisfy the
minimization principle:
\begin{equation}
\label{3.2}
{\delta \over \delta \alpha_i } \left( { \langle\Psi|{\cal
H}|\Psi\rangle \over \langle\Psi|\Psi\rangle} \right) =0\hspace{20mm}
i=1,N
\end{equation}
This leads to the system of CI equations
\begin{equation}
\label{3.3}
\sum_{j=1}^{N}{\cal H}_{ij} \alpha_j=E \sum_{j=1}^{N} {\cal O}_{ij}
\alpha_j
\hspace{20mm} i=1,N
\end{equation}
where $ E= \langle\Psi|{\cal H}|\Psi\rangle / \langle\Psi|\Psi\rangle$
is the energy of the system in the $|\Psi\rangle$ state , ${\cal
H}_{ij}= \langle \Psi_i| {\cal H}|\Psi_j\rangle$ are the matrix
elements of the Hamiltonian in the basis of Slater determinants
$\{|\Psi_i\rangle, i=1,N\}$, and ${\cal
O}_{ij}=\langle\Psi_i|\Psi_j\rangle$ are the overlap matrix elements
of the Slater determinants (which are not necessarily orthogonal). The
CI solution is easily found by solving the linear system of equations
(\ref{3.3}), once the basis of Slater determinants $\{|\Psi_i\rangle,
i=1,N\}$ is chosen.  If we denote by $\phi_p^{(n)}(i,\sigma)$ the
$p=1,...,N_e$ one-electron occupied orbitals of the Slater determinant
$|\Psi_n\rangle$, these matrix elements are given by:
\begin{equation}
\label{3.3b1}
{\cal O}_{nm}= \left|
\begin{array}[c]{ccc}
\beta_{1,1}^{nm} & ... & \beta_{1,N_e}^{nm} \\ \vdots & & \vdots\\
\beta_{N_e,1}^{nm} &... &\beta_{N_e,N_e}^{nm} \\
\end{array}
\right|
\end{equation}
The matrix elements of the Hamiltonian (\ref{1.104}) can be written as:
\begin{equation}
\label{3.3b2}
{\cal H}_{nm}= -t \cdot {\cal T}_{nm}+ U \sum_{i}^{}{\cal V}_{nm}(i)
\end{equation}
where the expectation values of the hopping and on-site interaction
terms are:
$$
{\cal T}_{nm}=\sum_{p=1}^{N} \left|
\begin{array}[c]{ccccc}
\beta_{1,1}^{nm}&...&t_{1,p}^{nm}&...&\beta_{1,N_e}^{nm}\\ \vdots &
&\vdots & & \vdots\\
\beta_{N_e,1}^{nm}&...&t_{N_e,p}^{nm}&...&\beta_{N_e,N_e}^{nm} \\
\end{array}
\right|
$$
and
$$
{\cal V}_{nm}(i)\!\!=\!\!\sum_{p_1 \neq p_2}^{} \!  \left|
\begin{array}[c]{ccccccc}
\beta_{1,1}^{nm} \!\!&...\!\!&u_{1,p_1}^{nm}(i)\!\!& ...
\!\!&d_{1,p_2}^{nm}(i)\!\!&...\!\! &\beta_{1,N_e}^{nm} \\ \vdots & &
\vdots & & \vdots & & \vdots\\
\beta_{N_e,1}^{nm}\!\!&...\!\!&u_{N_e,p_1}^{nm}(i)\!\!& ...
\!\!&d_{N_e,p_2}^{nm}(i)\!\!&...\!\! & \beta_{N_e,N_e}^{nm}\\
\end{array}
\right|_{.}
$$
Here,
$$
\beta^{nm}_{ph}= \sum_{{i\sigma}}^{}
\phi^{(n)*}_{h}(i,\sigma)\phi^{(m)}_{p}(i,\sigma),
$$
$$
t^{nm}_{p_1,p_2}= \sum_{\langle i, j \rangle
\atop\alpha\beta}^{}\left( \phi^{(n)*}_{p_1}(i,\alpha)
T^{ij}_{\alpha\beta} \phi^{(m)}_{p_2}(j,\beta) + h.c. \right),
$$
$$
u^{nm}_{p_1,p_2}(i)=\phi^{(n)*}_{p_2}(i\uparrow)\phi^{(m)}_{p_1}(i\uparrow), 
$$
and
$$
d^{nm}_{p_1,p_2}(i)=\phi^{(n)*}_{p_2}(i\downarrow)
\phi^{(m)}_{p_1}(i\downarrow).
$$

We now consider the specific choice of the Slater determinant basis
$\{|\Psi_i\rangle, i=1,N\}$.  Strictly speaking, one may choose an
optimized basis of Slater determinants from the general variational
principle:
\begin{equation}
\label{3.4}
{\delta \over \delta \phi^{(n)}_p(i,\sigma) } \left( {
 \langle\Psi|{\cal H}|\Psi\rangle \over \langle\Psi|\Psi\rangle}
 \right) =0\hspace{5mm} n=1,N; p=1,N_e
\end{equation}
However, implementation of this full trial-function minimization
scheme (also known as a multi-reference self-consistent mean-field
approach \cite{Fulde}) is numerically cumbersome even for medium-sized
systems. Instead, we select the Slater determinant basis
$\{|\Psi_i\rangle, i=1,N\}$ from the set of broken symmetry,
Unrestricted Hartree-Fock wavefunctions (\ref{2.3}), their symmetry
related partners and their excitations. Clearly, (\ref{2.3}) satisfies
(\ref{3.4}) by itself, provided that the $\alpha$ coefficients
corresponding to the other Slater determinants in Eq.(\ref{3.1}) are
set to zero (see Eq. (\ref{2.1})).  Since this Unrestricted HF
wavefunction is not translationally invariant (the doping hole is
always localized somewhere on the lattice), we can restore the
translational invariance of the CI ground-state wavefunction by also
including in the basis of Slater determinants all the possible lattice
translations of this Unrestricted HF wavefunction. 
 Furthermore, if the self-consistent configuration is
not rotationally-invariant (e.g. a meron-antimeron pair), all possible
rotations must be performed as well. By rotation we mean changing the
relative position of the meron and antimeron while keeping their
center of mass fixed. 

Clearly, all the translated 
HF Slater determinants lead to the same HF
ground-state energy $\langle \Psi_n| {\cal H} | \Psi_n\rangle =
E_{GS}$ as defined by Eq. (\ref{2.8}). The CI method lifts the
degeneracy between states with the hole-induced configuration
localized at different sites (see Eq. (\ref{3.3})), 
thereby restoring translational 
invariance. We may identify the lowering in the total energy due to
the lifting of this degeneracy as quantum mechanical kinetic energy of
deconfinement which the doping-induced configuration saves through
hopping along the lattice. In addition, quantum fluctuations in the
internal structure of a magnetic soliton can be incorporated by
including the lowest order excited state configurations of the static
Hartree-Fock energy spectrum. Such wavefunctions are given by
$a^{\dagger}_p a_h |\Psi\rangle$, where $p > N_e$ labels an excited
particle state and $h \le N_e$ labels the hole which is left behind
(see Eq. (\ref{2.3})). Once again, all possible translations (and
non-trivial rotations) of this
``excited'' configurations must be included in the full CI
wavefunction. These additions can describe changes in the ``shape'' of
the soliton as it undergoes quantum mechanical motion along the plane.

The CI method is described in more detail in Reference 3, where
it is used to study the 1D Hubbard chain in order to gauge its
accuracy by comparing its results with the exact Bethe ansatz solution. 
 We showed that the CI  method recaptures the
essential physical features  of the exact
solution of the 1D Hubbard chain, such as spin-charge separation, 
 as well as leading to remarkable
agreement of ground state energies of doped chains for all values
of $U/t$.
  The main difference between the 1D case and the
2D case is the computation time required. The computation time for one
matrix element ${\cal H}_{nm}$ scales roughly like $N^9$, where $N$ is
the number of sites. The number of configurations included in the CI
set scales as $N! / N_s!(N-N_s)!$ when $N_s$ solitons are present.
For both an $N$-site chain and an $N\times N$ lattice, the HF ``bulk''
limit is reached for $N \ge 10$. In the 1D case \cite{prev} we used chains with
$N=10-25$, and numerical calculations can be easily
performed. However, in 2D the smallest acceptable system has 100
sites, leading to an enormous increase in the computation
time. Nevertheless, our sample of results in 2D suggest a simple and
clear physical picture which we describe below.

\subsection{\label{ssec5.4.1}Spin-bag Dissociation: \\ Spin-Charge
Separation in 2D }

The charged spin-bag carries a spin of 1/2.  Let
$|\Psi_{+}\rangle, |\Psi_{-}\rangle$ be the HF determinants for
the spin-bag centered at any two nearest neighbor sites, respectively, and
let $\hat{S}_z= \sum_{i}^{}\hat{S}_z(i)={1\over2} \sum_{i,\sigma}^{}
\sigma c^{\dagger}_{i\sigma} c_{i\sigma}$ be the total spin operator
in the $z$-direction.  Then, $\hat{S}_z|\Psi_{+}\rangle={1\over2}
|\Psi_{+}\rangle$ while $\hat{S}_z |\Psi_{-}\rangle=-{1\over 2}
|\Psi_{-}\rangle$ (or viceversa), since moving the center of the 
spin-bag by one site leads to a flip of its total spin (see
Fig. 3(a)). Consequently,  $ \langle
\Psi_{-}|\Psi_{+}\rangle=0$. Since the Hubbard Hamiltonian
commutes with $\hat{S}_z$, it follows that $\langle \Psi_{-}|{\cal
H}|\Psi_{+}\rangle=0$. From the CI equation (\ref{3.3}) we conclude
that there is no mixing between states with different total spin. 
As a result, it is enough to include in the CI set
only those configurations with the spin bag localized on the same
magnetic sublattice.  Let us denote by $|\Psi_{(0,0)}\rangle$ the
initial static Hartree-Fock configuration, and by
$|\Psi_{(n,m)}\rangle$ the configuration obtained through its
translation by $n$ sites in the x-direction and $m$ sites in the
y-direction (cyclic boundary conditions are imposed). The condition
that only configurations on the same sublattice are included means
that $n+m$ must be an even number, and the cyclic boundary conditions
mean that $0\le n\le N-1, 0\le m\le N-1$, for a $N \times N$
lattice. As explained in detail in the 1D analysis, \cite{prev} mixing
configurations with the charged spin-bag localized at different sites
and then subtracting out the contribution of the undoped AFM
background allows us to calculate the dispersion band of the spin-bag
itself:
\begin{equation}
\label{3.40}
E_{sb}(\vec{k})=E(\vec{k},N) - N^2 e_{GS}.
\end{equation}
Here, the total energy of the lattice with the spin-bag  
$$
E(\vec{k},N)= { \langle \Psi_{\vec{k}}| {\cal H} |
\Psi_{\vec{k}}\rangle \over \langle \Psi_{\vec{k}}|
\Psi_{\vec{k}}\rangle}
$$
and the CI wave-function
$$
|\Psi_{\vec{k}}\rangle = \sum_{(n,m)}^{} \exp{\left( i (k_x n +k_y
|m)a\right)} \Psi_{(n,m)}\rangle
$$
are the solutions of the CI equations (\ref{3.3}). The finite size of
the lattice and cyclic boundary conditions restricts
 the calculation to $\vec{k}$-points  of the form
$\vec{k}= {2\pi\over Na} \left( \alpha \vec{e}_x +\beta\vec{e}_y\right)$, where
$(\alpha,\beta)$ is any pair of integer numbers. As usual, only
$\vec{k}$-points inside the first Brillouin zone need to be considered.

An analysis of the dependence of the spin-bag dispersion relation
$E_{sb}(\vec{k})$ on the size $N\times N$ of the lattice is shown in
Fig. \ref{fig3.20}, for the conventional Hubbard model (upper panel)
and spin-flux model (lower panel), and $U/t=5$. We used 6x6, 8x8,
10x10 and (only for the spin-flux model) 12x12 lattices.  The
dispersion relation is plotted along lines of high symmetry of the
full Brillouin zone. For comparison, we also show the excitation
energy $E_{sb}^{HF}$ obtained in the static HFA as a full line. For
both models, we see that the spin-bag dispersion band is almost
converged, even though we used quite small lattices. The convergence
is somewhat slower in the spin-flux case, as seen most clearly at the
(0,0) point. Although the values obtained from the four lattices all
differ at (0,0), the extremum values correspond to the 6x6 and the 8x8
lattices, while the values for the 10x10 and 12x12 lattices are
indistinguishable. We conclude that the fit (\ref{3.40}) is
legitimate.

From Fig. \ref{fig3.20} we also see that the dispersion relations for
the spin-bag in the two different models are very different. The
dispersion relations over the full 2D Brillouin zone are shown in
Fig. \ref{fig3.21}, and they are seen to mimic the electronic
dispersion relation of the underlying undoped AFM background, shown in
Fig. \ref{fig2.13}. This is consistent with the quasi-particle nature
of this charged spin-1/2 spin-bag. In the conventional Hubbard model, the
undoped AFM background has a large nested Fermi surface along the $(0,\pi)$ to
$(\pi,0)$ line, and it is exactly along this line that the spin-bag
dispersion band has a minimum. Similarly, the lowest energy of the
spin-bag of the spin-flux model is at $(\pi/2, \pi/2)$, corresponding
to the Fermi points of the underlying undoped spin-flux AFM
background.

The extra kinetic energy $E_{sb}({ \pi \over 2} ,{ \pi \over 2})-
E_{sb}^{HF}$ saved by the spin-bag through quantum hopping is $0.37 t$
in the conventional model and $0.56t$ in the spin-flux model (for
$U/t=5$). Since the spin-bag is confined to  one magnetic
sublattice, it must tunnel two lattice constants to the next allowed
site. Consequently, the energy gained through hopping (of order
$t^2/U$) is small. This is displayed, for the spin-bag of the
spin-flux model, in Fig. \ref{fig3.24}, where we plot the lowering in
kinetic energy of the deconfined spin-bag $E_{sb}({ \pi \over 2} ,{
\pi \over 2})-E_{sb}^{HF}$ as a function of $t/U$. A similar
dependence for the spin-bag of the conventional Hubbard model is
presented elsewhere.  \cite{CIpaper} As in the 1D case, we conclude
that the spin-bag in 2D is a rather immobile quasiparticle-like
excitation.

In the 1D model it is energetically favorable for the immobile
 spin-bag to decay
 into a charged bosonic domain wall and a neutral fermionic domain
 wall, resulting in spin-charge separation. \cite{prev} The analog of the 1D
 charged bosonic domain wall is the 2D charged bosonic meron-vortex of
 the spin-flux model. If the spin-bag decays into a charged
 meron-vortex, a magnetic antivortex must also be created for
 topological reasons.  Unlike the pair of domain walls in the 1D case,
 the vortex-antivortex pair is tightly bound by a topological binding
 potential that increases as the logarithm of the vortex-antivortex
 separation. Therefore, we expect that the doping charge is shared
 between the two magnetic vortices. One technical problem for testing
 this hypothesis is that such a configuration ( a vortex-antivortex
 pair sharing one doping hole) is not self-consistent at the static
 Hartree-Fock level. In the static approximation we require two doping
 holes to stabilize two vortex cores and create a meron-antimeron
 pair.  We can, however, construct a  trial
 wavefunction to describe the singly-charged vortex-antivortex pair,
 by adding one electron in the first empty state of the
 self-consistent doubly-charged meron-antimeron configuration. The
 first empty levels of the meron-antimeron pair are the localized
 levels bound in the vortex cores, two for each vortex (see
 Fig. 4(b)). Because of degeneracy between the two lower
 discrete levels of the pair, we have in fact two distinct trial
 wave-functions, obtained by adding one electron in either of these
 two lower localized gap electronic states of a self-consistent
 meron-antimeron pair.  These wavefunctions are not invariant to
 rotations ( see Fig. \ref{fig2.25}).  Therefore we must include in
 the CI set of Slater determinants the configurations obtained through
 $\pi/2$ rotations of the vortex-antivortex pair about its fixed
 center of mass in addition to translated configurations.  As a
 result, we have a total of $8N^2$ configurations describing the
 singly charged vortex-antivortex pair localized at all possible sites
 with all possible orientations about the center of mass.

We performed this CI analysis for a $10 \times 10$ lattice and
$U/t=5$. The HF energy of a  simple static
spin-bag is $-0.82 t$ (measured with respect to the HF energy
of the undoped AFM  background, equal to $-76.76 t$). The energy of the static
singly-charged vortex-antivortex pair is $-0.23 t$. Thus, we see that
because this singly-charged pair trial wavefunction is not
self-consistent, in the static case this configuration is
energetically much more costly than the self-consistent spin-bag
configuration. However, if we allow for quantum motion of these
configurations, the situation changes dramatically. Performing the CI
analysis for the set of all possible translated spin-bag
configurations, we find that the energy of the spin-bag is lowered to
$-1.24 t$. Performing the CI analysis for the set of all translated
and rotated singly-charged vortex-antivortex pairs we find that this
configuration's energy is lowered to $ -2.18 t$. This shows that the
vortex-antivortex pair has lowered its translational and rotational
kinetic energy by almost $2t$, thereby becoming the low-energy charge
carrier. This large number is not surprising, since unlike the
spin-bag, the vortex-antivortex pair is not constrained to motion on
one magnetic sublattice. As a result, such configurations lower their
kinetic energy by an amount on the scale of $t$, as opposed to $t^2/U$
for the spin-bag configuration.  For larger $U/t$ values this effect is
even more pronounced.

We conclude that these results strongly support the hypothesis of
spin-bag dissociation into a much more mobile singly-charged
vortex-antivortex pair, analogous to the 1D spin-bag dissociation into
a pair of a charged bosonic domain-wall and a neutral fermionic
domain-wall. \cite{prev} Unlike in the 1D case, however, we do not
have distinct charge and spin carriers for the composite
excitation. Instead the spin and the charge are shared equally between
the vortex and the antivortex. If, on the other hand, there was a
mechanism whereby the vortices became unbound, complete spin-charge
separation could occur, in which one vortex traps the hole (and is
therefore a charged meron) and the other vortex carries the spin, in a
lotus-flower\cite{MB1,MB2} (or undoped magnetic meron)
configurations. In the absence of the corresponding self-consistent
static HF configuration we are not able to settle this question. At
very low doping, the strong vortex-antivortex topological attraction
binds the spin and charge together. This is different from the 1D
case, where the absence of long-range interactions between the
domain-walls allow for a complete spin-charge separation at any doping
and even at zero temperature.

This scenario opens a new avenue for research into how the system
evolves with doping. If each hole is dressed into a singly-charged
vortex-antivortex pair, when two such pairs overlap it is possible
that both doping charges move to the same pair, creating a
meron-antimeron pair of charged bosons. Such pre-formed charge pairs
may condense into a superconducting state at low temperatures.  The
other uncharged vortex-antivortex pair may either collapse and
disappear (this is likely to happen at low-temperatures) or remain as
a magnetic excitation of the system (at higher temperatures),
mediating the destruction of the long-range AFM order, the
renormalization of the spin-wave spectrum and the opening of the spin
pseudogap.

\subsection{\label{ssec5.4.2}D-wave pairing of charge carriers}

From the static HF analysis we found that the most stable static
self-consistent configuration with two doping holes added to the AFM
background of the spin-flux model is the meron-antimeron pair, for $3
< U/t < 8$. At larger $U/t$, two charged spin-bags become more stable,
in the static HF approximation. This is in close analogy to the
prediction that the spin-bag is energetically more favorable than the
static charged domain-wall for $U/t > 6.5$, in the HFA of the 1D
Hubbard model. \cite{prev} However, in the 1D case the charged
domain-wall is considerably more mobile than the charged spin bag,
gaining a kinetic energy of the order of $t$ as opposed to $t^2/U$
energy gained by the spin-bag. As a result, when this kinetic energy
of deconfinement is taken into account within the CI method, the
charged domain-wall is found to be the relevant charged excitation for
all values of $U/t$. A similar picture emerges in the 2D case, because
the meron-vortices are much more mobile than the spin-bags.

For the 2D system, we have shown that the charged spin-bag has very
similar behavior to the 1D charged spin-bag. The analog of the 1D
charged bosonic domain-wall is the 2D charged bosonic meron-vortex.
We now consider the properties of the doubly-charged meron-antimeron
pair. All the numerical results quoted in the rest of this section
refer to a meron-antimeron pair on a 10x10 lattice, in the spin-flux
model with $U/t=5$.

As already discussed, the meron-antimeron pair is not rotationally
invariant. We can find the rotational kinetic energy saved by the pair
as it rotates about its center of mass. In the present case, only 4
configurations need to be included, corresponding to the four possible
self-consistent arrangements of the meron and antimeron about their
fixed center of mass (see Fig. \ref{fig3.28}).  Simple rotation by
$\pi/2$ of the one-particle orbitals $\phi_p(i,\sigma)$ 
about the center of mass is not,
however, sufficient to generate the rotated configurations. First of
all, the $\pi/2$ rotation also changes the spin-flux
parameterization. If the spin-flux of the initial configuration is
$T^{12}=-1, T^{23}=T^{34}=T^{41}=1$, a $\pi/2$ rotation leads to a
state corresponding to the rotated configuration $T^{12}=1, T^{23}=-1,
T^{34}=T^{41}=1$.  Thus, following the $\pi/2$ rotation, a unitary
transformation must be performed in order to restore the initial
spin-flux parameterization. For the case cited above, this simply
implies the change in the one-particle orbitals
$\phi_p(i_x,i_y,\sigma) \rightarrow - \phi_p(i_x,i_y,\sigma)$ for all
sites $(i_x,i_y)$ which are a type '2' site of the unit cell, in other
words sites with even $i_x$ and odd $i_y$ (also, see
Fig. \ref{fig1.55}). The second observation is that the rotation by
$\pi/2$ also changes (flips) all the spins of the AFM background
surrounding the pair. Thus, an extra $\pi$ rotation about an axis
perpendicular to the lattice plane is necessary to restore the
alignment of the AFM background. Following these transformations it is
straightforward to generate the Slater determinants $|\Psi_2\rangle,
|\Psi_3\rangle$ and $|\Psi_4\rangle$ corresponding to the
meron-antimeron pairs rotated by ${\pi\over 2}, \pi$ and $3\pi \over
2$ from the initial self-consistent HF meron-antimeron pair described
by $|\Psi_1\rangle$.  The CI method can be used to find the rotational
energy saved by superposing these rotated meron-antimeron
configurations.  The lowest CI energy found is $0.46t$ below the
energy of the static pair, and corresponds to d-wave symmetry. By this
we mean that the coefficients $\alpha_i$ multiplying the 4 rotated
states in the CI wave-function $|\Psi\rangle = \sum_{i=1}^{4} \alpha_i
|\Psi_i\rangle$ satisfy the condition
$\alpha_1=-\alpha_2=\alpha_3=-\alpha_4$.

Translation of a pair over the whole lattice can also be
investigated. Since the pair does not carry any spin, all possible
translations must be included (there is no restriction to same
magnetic sublattice configurations). This leads to a total of $N^2$
possible configurations for a $N\times N$ lattice. Again, when various
configurations are generated from the initial self-consistent HF
meron-antimeron state $ |\Psi_1\rangle$, care must be taken to
preserve the same spin-flux parameterization and the same AFM
background orientation. This can be achieved performing
transformations similar to the ones described above. As a result of
performing the CI method on the set of translated states, we find the
dispersion relation of the (un-rotated) meron-antimeron pair. This is
shown in Fig. \ref{fig3.25}. In this plot we show the total energy of
the lattice with the moving meron-antimeron pair, as a function of the
total momentum of the pair. Quantum hopping of the meron-antimeron
pair lowers its total energy by an extra $1.29t$. Two other
interesting features are observed in Fig. \ref{fig3.25}. The first one
is that the dispersion relation of this rigidly polarized pair is not
invariant to rotations by $\pi/2$, as expected.  More important is
that the minima of the dispersion relation occur at the $(\pi,\pi)$
points. Since the momentum of the pair is twice the momentum of either
the meron or the antimeron, this is consistent with the fact that in
their lowest energy state, both the meron and the antimeron have
momenta of $(\pi/2, \pi/2)$, in the spin-flux model. The doubling of
the size of the Brillouin zone is also a direct consequence of this
doubling of total momentum ( for comparison with undoped dispersion
relation, see Fig. \ref{fig2.13}).

However, to obtain the true energy of the charged pair, we must mix
all $4N^2$ rotated and translated meron-antimeron configurations. All
have the same static HF energy and are equally important in the CI
method.  Let $|\Psi_0(0,0)\rangle$ denote the initial self-consistent
static Hartree-Fock meron-antimeron configuration, and $|
\Psi_{\theta}(n,m)\rangle$ denote the configuration obtained through
translation by $n$ sites in the x-direction and $m$ sites in the
y-direction, as well as a rotation by an angle of $\theta{\pi \over
2}$ of the pair about its center of mass. Here $0 \le \theta \le 3$
and $0 \le n \le N-1 , 0 \le m \le N-1$ (cyclic boundary conditions
are imposed).  The CI wavefunctions are then given by:
\begin{equation}
\label{3.100}
|\Psi_{\vec{k}}\rangle = \sum_{(n,m)}^{} e^{i \left( k_x n + k_y
 m\right) a } \left(\sum_{\theta=0}^{3} \alpha_{\theta}
 |\Psi_{\theta}(n,m)\rangle \right)
\end{equation}

The dispersion relation
$$E_{pair}(\vec{k})= {\langle \Psi_{\vec{k}}| {\cal H} |
\Psi_{\vec{k}}\rangle \over \langle \Psi_{\vec{k}}|
\Psi_{\vec{k}}\rangle } - E_{pair}^{HF}$$ obtained from this complete
set is shown in Fig. \ref{fig3.26}. The reference point is the HF
energy of the self-consistent meron-antimeron pair
$E_{pair}^{HF}=-78.52t$.  Thus, we see that the total kinetic energy
saved by the freely moving and rotating meron-antimeron pair is
$1.75t$, when the total momentum of the pair is $(\pi,\pi)$. This
energy equals the sum $0.46t +1.29t$ of rotational and translational
kinetic energies found before (the number of significant figures
indicates the estimated accuracy of the computational method).  The
rotational invariance of the dispersion band is also restored. Besides
the absolute minima about the $(\pi,\pi)$ points, there is a more
shallow minimum region about the $(0,0)$ point.

The rotational symmetry of the meron-antimeron pair wavefunction,
defined by the coefficients $\left(\alpha_{\theta}\right)$, is a
function of the total momentum carried by the pair, as shown in
Fig. \ref{fig3.27}. The absolute minima points $(\pi,\pi)$ and the
area around them correspond to pairs with d-wave symmetry. By this, we
mean that the coefficients $\alpha_{\theta}$ have the form
\begin{equation}
\label{3.101}
\alpha_{\theta}= \exp{\left( i J \theta {\pi \over 2}\right)} \alpha_0
\end{equation}
with $J=2$, i.e. $\alpha_0=-\alpha_1=\alpha_2=-\alpha_3$.  The core
area, about the local minimum $(0,0)$ point, corresponds to s-wave
symmetry. In this region the coefficients $\left (
\alpha_{\theta}\right)$ again satisfy Eq. (\ref{3.101}), but for
$J=0$, i.e. $\alpha_0=\alpha_1=\alpha_2=\alpha_3$.  The intermediary
area appears to be a mixture of different $J$ values. A simple
decomposition of the form (\ref{3.101}) is no longer
possible. Instead, a sum of such terms corresponding to different $J$
values is required. Since we only have rotations by $\pi/2$, a unique
identification of the composite symmetry is not possible.  Moreover,
the energy of the states in this intermediary area is at the top of
the dispersion band. In order to find the correct CI states for
energies well above the static HF value (i.e. larger than zero, in
this case) we must add to the CI set the first set of excited HF
states. For a meron-antimeron pair, excitation of an electron from the
valence band onto the empty localized levels inside the Mott-Hubbard
gap costs about 1.5t of energy, for $U/t=5$, so such states should
contribute significantly in the CI states with positive energies and
modify their dispersion and symmetry (for this reason, we do not show
the upper three high-energy bands in Fig. \ref{fig3.26}).
Consequently, both the energy and the symmetry of the states in the
intermediary area may be modified from the ones shown in
Fig. \ref{fig3.26}.  However, the minima at $(\pi,\pi)$ and $(0,0)$
are at energies well below zero. Their energies and rotational
symmetry are unaffected by additions of higher energy configurations
to the CI set.

The fact that we obtain two distinct minima is not very surprising. As
argued before, we expect that individual merons and antimerons are
created with momenta of $(\pm \pi/2,\pm \pi/2)$. As a result, two
different couplings are possible. A $(\pi/2,\pi/2$) meron can pair
with a $(\pi/2,\pi/2$) antimeron, creating a pair of total momentum
$(\pi,\pi)$. This is the most stable coupling, leading to the lowest
possible energy of $-1.75t$ below the static HF energy. This pair has
d-wave symmetry. The second possible coupling is between a
$(\pi/2,\pi/2$) meron and a $(-\pi/2,-\pi/2$) antimeron. This pair has
a total momentum of $(0,0)$, and s-wave symmetry. However, this
coupling is less strong. For the $U/t=5$ case considered, the energy
of the s-wave $(0,0)$ pair is $1.28t$ above the energy of the d-wave
$(\pi,\pi)$ pair. The existence of both d-wave and s-wave pairing, and
the dominance of the d-wave pairing, have been established
experimentally for the high-T$_c$ cuprates. \cite{Harlin} We are not
aware of any other microscopic theory that predicts the two types of
pairing to appear in different regions of the Brillouin zone.

The total kinetic energy saved by the meron-antimeron pair through
quantum hopping and rotation is of order $t$, as expected, since the
pair is not restricted to one magnetic sublattice and tunneling is not
required for motion.  Consequently, we expect that the energy saved by
the meron-antimeron pair for larger values of $U/t$ is comparably
large. On the other hand, the energy saved by the spin bag through
tunneling motion scales like $t^2/U$. In fact, we argued that a
spin-bag may dissociate into a singly-charged vortex-antivortex pair
in order to enhance its mobility. However, even if dissociation does
not occur, the kinetic energy saved by a pair of spin-bags is
significantly smaller than the kinetic energy saved by the
meron-antimeron pair. This shows that for $U/t=5$ the meron-antimeron
pair is even more favorable energetically than the HFA predicts and
suggests that the $U/t$ range where meron-antimeron pair formation
occurs may extend well beyond the $U/t=8$ limit found within the
HFA. \cite{MB3} In the 1D case, the range of stability of the charged
domain wall versus the charged spin-bag is extended (from the HF
prediction of $U/t=6.5$) to all $U/t$ range.\cite{prev} A numerical
analysis is needed to determine if the limit is extended to infinity
in the 2D case as well.

\section{Summary and Conclusions}

The Configuration Interaction Approximation incorporates crucial
quantum translational and rotational motion of the charge carriers,
which are absent in the static Hartree-Fock Approximation.  Given the
accuracy of the CI method in recapturing certain features of the exact
Bethe ansatz solution of the 1D Hubbard model, \cite{prev} we believe
that the CI method is likewise a very powerful tool for describing
effects beyond mean-field theory in 2D.  For 2D systems, numerical
calculations are much more time consuming.  However, our small sample
of results is quite suggestive of a simple physical picture. In direct
analogy with the 1D results, we find that the bosonic charged
meron-vortices are much more mobile than the fermionic charged
spin-bags. The extra kinetic energy gained by the meron-vortices very
likely extends their region of stability beyond the $U/t=8$ limit
suggested in the HFA. \cite{MB3} There are also strong indications
that a charged spin-bag is unstable to decay into a singly-charged
vortex-antivortex pair, analogous to the spin-charge separation in the
1D case.  Nucleation of such pairs of vortices with doping is expected
to further influence the magnetic behavior of the cuprates. The bound
state and the unbound continuum states of the singly charged
meron-antimeron pair may account for the anomalous ``quasiparticle''
spectral widths observed on angle-resolved photo-emission experiments.

The symmetry of the meron-antimeron pairs emerges very clearly from
the CI treatment. We find two regions of stability of the
meron-antimeron pair. Pairs with total momentum of $(\pi,\pi)$ have
d-wave symmetry, and are the most stable. Pairs with total momentum
$(0,0)$ have s-wave symmetry and have a smaller gap. Thus, we find
that different pairing appears in different regions of the Brillouin
zone.  These results appear to have a direct bearing on numerous
experiments, which show a mixture of strong d-wave component and a
smaller s-wave component in the superconducting state of the
cuprates. \cite{Harlin}

Many other features of our model are in agreement with experimentally
observed properties of the cuprate superconductors. Nucleation of
magnetic vortices with doping explains a variety of magnetic
properties, starting with complete destruction of the long-range AFM
order for very low doping concentration. As we can see from
Fig. 5(a) a tightly-bound meron-antimeron pairs disturbs the
long-range AFM ordering of most of the spins on the 10x10 lattice. For
very low dopings, these pairs are far from each other, and there are
many spins on the plane whose orientations are not affected by any
pair. Thus, most of the spins maintain the long range AFM
order. However, as the doping increases to about $5 \%$ (which is
roughly equivalent to having two meron-antimeron pairs on the 10x10
lattice) the areas occupied by each meron-antimeron pairs start to
overlap with those occupied by the neighboring pairs. At this doping
the orientation of all the spins on the CuO$_2$ planes is affected by
at least one pair of vortices, and therefore the LRO is severely
disrupted. The local ordering, however, is still AFM. This picture
explains the extremely low doping necessary for the disappearance of
LR AFM order, as well as the fact that the spin correlation length is
basically equal to the average distance between holes (vortices) and
does not depend strongly on the temperature.\cite{Birge} Each hole
carries its vortex with it, and the spins in each vortex are
correlated with each other. The correlation length is thus roughly
equal to the average size of the vortex which is defined by the
average inter-vortex (inter-hole) distance.  The nucleation of
magnetic vortices also explains the split of the $(\pi,\pi)$ AFM Bragg
peak into the four incommensurate peaks whose positions shift with
doping,\cite{vakh} as observed in LaCuO and, more recently, in
YBaCuO.\cite{arai} The form factor of any given vortex already gives
rise to an apparent splitting of the neutron scattering peak.  As
demonstrated in Ref. 9, even at the mean-field level we
recapture the neutron scattering data using a random distribution of
meron-vortices.

Optical behavior of the cuprates is also explained naturally using our
model. Two features develop in the optical absorption spectra with
doping: a broad mid-infrared temperature-independent absorption band,
and a strongly temperature-dependent low-frequency Drude
tail. \cite{optics} In our model the broad mid-infrared band is
related to excitation of electrons from the valence band onto the
empty levels bound in the vortex cores, \cite{MB3} which are localized
deep inside the Mott-Hubbard gap (see Fig. 4(b)). The number of
localized levels scales with the number of vortices, and inter-vortex
interactions lead to their splitting into a broad band.  As such, this
mechanism is similar to the one leading to a broad mid-infrared
absorption band in polyacetylene with doping. \cite{Tanaka} The
polyacetylene band is due to electronic excitations inside the cores
of the polyacetylene domain-wall solitons,\cite{polrev} which are the
topological analogues of meron-vortices.\cite{MB1,MB2} Another strong
argument in favor of this interpretation is provided by photoinduced
absorption experiments. \cite{foster} If the undoped parent compounds
are illuminated with intense visible light, they develop absorption
bands that resemble the mid-infrared bands of the doped
compounds. Similar behavior is observed in polyacetylene, and is
attributed to the nucleation of solitons by photoexcited electron-hole
pairs. \cite{Orenstein} The second component of the optical spectrum
is the Drude tail. It results from the response of the freely moving
charged vortices to the external electric field. The strong
temperature dependence of this tail is determined by the scattering
mechanism for merons (presumably due to interactions with other merons
and spin-waves). This interpretation is also supported by the fact
that the superconducting transition leaves the mid-infrared absorption
band unchanged.  Merons with internal electronic structure are still
present on the planes but pair condensation leads to a collapse of the
Drude tail into a $\delta(\omega)$ response.

As already discussed, nucleation of charged meron-vortices with doping
provides a microscopic basis for non-Fermi-liquid behavior, due to the
bosonic nature of the mobile charged excitations. The model also
predicts the existence of pre-formed pairs with d-wave symmetry, which
are thought to be responsible for the pseudo-gap
effects. \cite{pseudo} As the number of pairs of charged bosons
increases with doping and the temperature is lowered, the
meron-antimeron pairs Bose
condense and become coherent, leading to superconductivity. 
This mechanism naturally explains the puzzling scaling of the
superfluid density with doping $\rho_s \sim \delta$, in other words
with the number of holes, not of electrons. In our model, it is the
doping-induced positively charged meron-vortices that are the mobile charge
carriers. 
As a result, the density of preformed meron-antimeron  Cooper pairs is
obviously proportional to doping.  
Finally,
for large dopings ($\delta > 0.30-0.40$) the average vortex size
become extremely small and the very cores of the merons start to
overlap. In this limit the Mott-Hubbard gap is completely filled in by
the discrete levels, and the spin-flux state becomes energetically
unstable relative to a normal Fermi liquid. \cite{MB3}

It is noteworthy that all of the independent features described above
are in qualitative agreement with our model, which has essentially no
free or adjustable parameters. The choice of $U/t$ is fixed by the
experimentally measured size of the Mott-Hubbard charge transfer gap
at zero doping. More detailed comparisons with specific experiments
may require the incorporation of specific (smaller energy scale)
interactions which are not included in this simplest version of the
spin-flux Hamiltonian. A derivation of the explicit consequences of
this picture appears to be worthy. A more comprehensive and
quantitative e comparison with the experiments may prove quite
fruitful.

\section*{Acknowledgments}

M.B. acknowledges support from the Ontario Graduate Scholarship
Program and a fellowship from William F. McLean.  This work was
supported in part by the Natural Sciences and Engineering Research
Council of Canada.

\newpage

\begin{figure}
\centering
\parbox{0.45\textwidth}{\psfig{figure=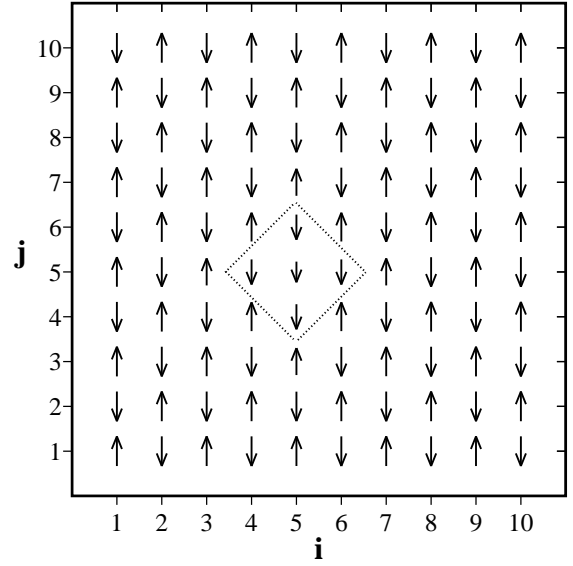,width=85mm}}
\parbox{0.45\textwidth}{\psfig{figure=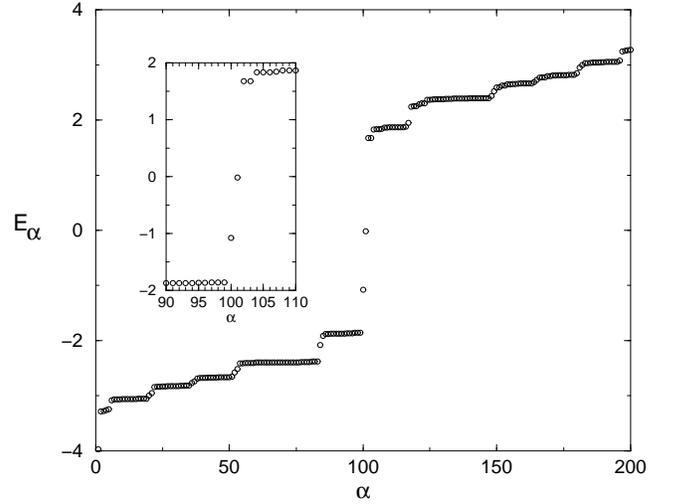,width=85mm}}
\caption{\label{fig2.16} (a) Self-consistent spin distribution of a
10x10 lattice with a spin bag centered at (5,5). The spin-bag has a
small ferromagnetic core, and the magnetic order is only locally
affected. (b) Electronic spectrum of a spin bag on a 10x10 lattice,
for $U/t=5$ in the spin-flux model.  Eigenenergies $E_{\alpha}$ are
plotted as a function of $\alpha=1,200(=2N^2)$. Only the first
$N^2-1=99$ states are occupied. There are two empty bound discrete
levels deep into the Mott-Hubbard gap ($\alpha=100, 101$), one of
which comes from the valence band of the undoped AFM compound (see
inset). There is also an occupied discrete level below the valence
band ($\alpha=1$). The valence band is spin paired, since it has an
even number of levels. Thus, the total spin of the spin-bag comes from
the discrete occupied level. The spin-bag is a charged, spin-${1\over
2}$ fermion.}
\end{figure}

\begin{figure}
\centering
\parbox{0.45\textwidth}{\psfig{figure=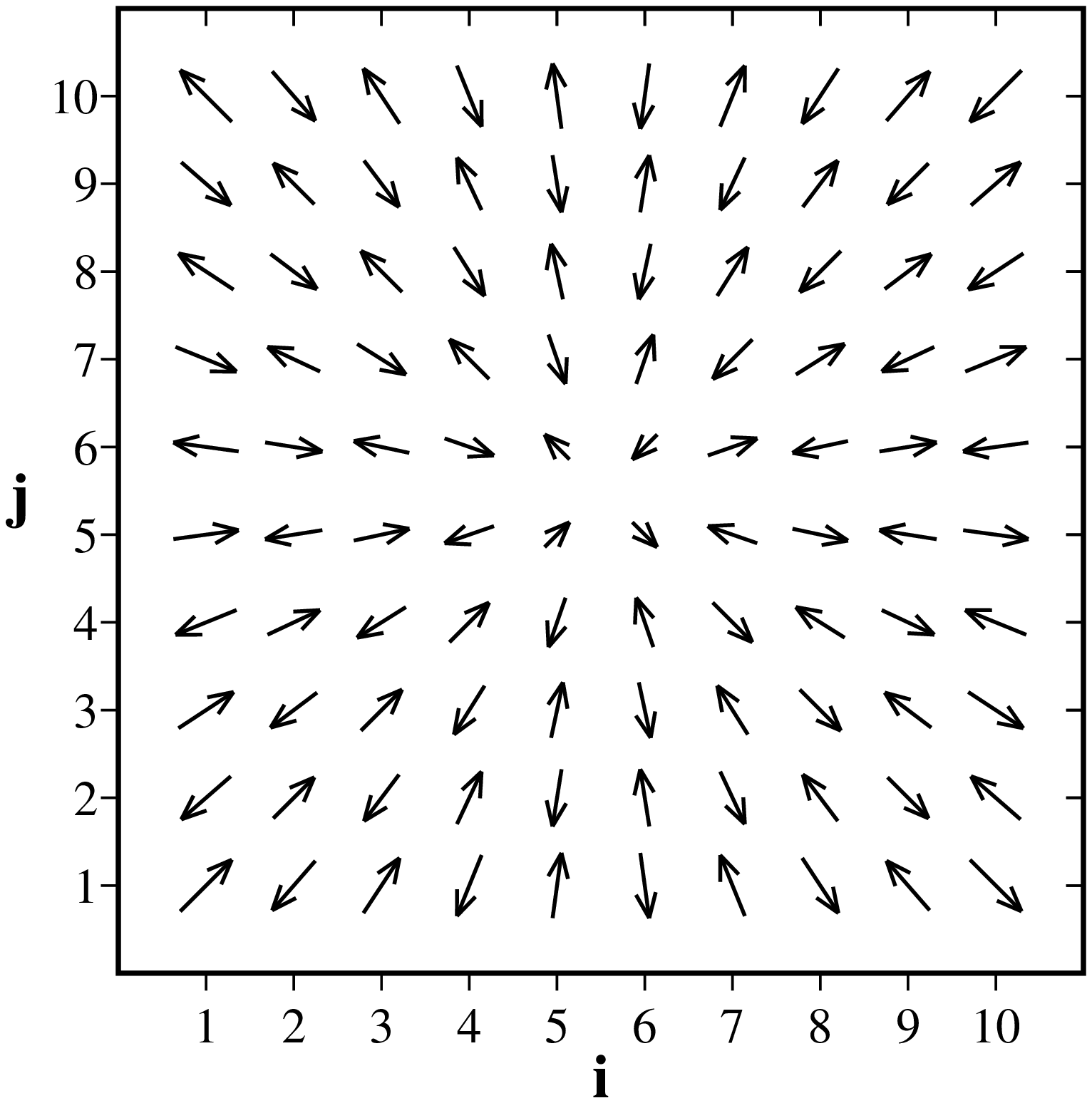,width=85mm}}
\parbox{0.45\textwidth}{\psfig{figure=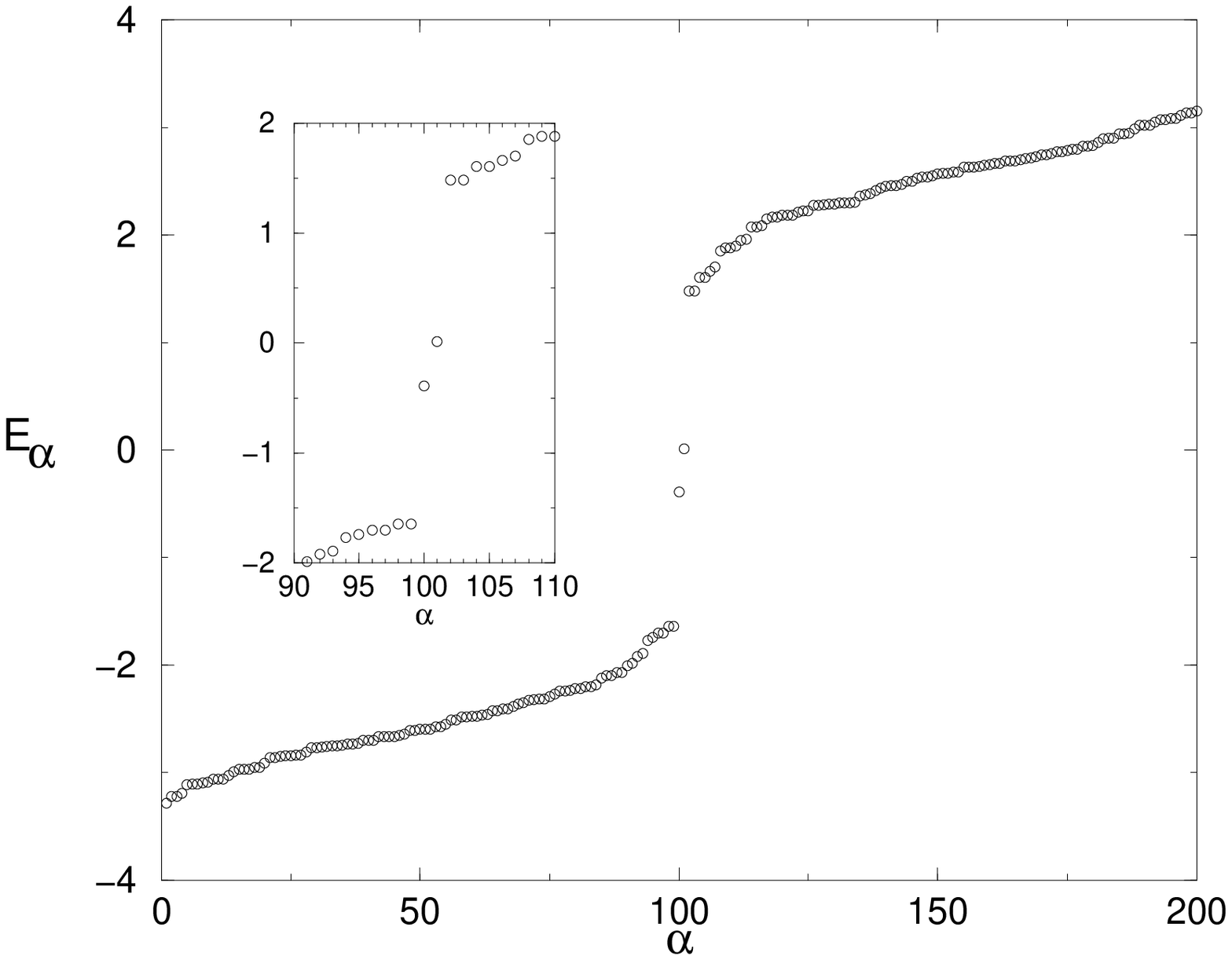,width=85mm}}
\caption{\label{fig2.22} (a).  Self-consistent spin distribution of a
10x10 lattice with a meron-vortex in the spin-flux phase. The core of
the meron is localized in the center of a plaquette, in the spin-flux
phase (in the conventional phase, the core of the meron-vortex is
localized at a site). This excitation has a topological winding number
1, since the spins on either sublattice rotate by $2\pi$ on any curve
surrounding the core. The magnitude of the staggered magnetic moments
is slightly diminished near the vortex core but is equal to that of
the undoped AFM background far from the core.  (b). Electronic
spectrum of a meron-vortex on a 10x10 lattice, for $U/t=5$, in the
presence of the spin-flux. Eigenenergies $E_{\alpha}$ are plotted as a
function of $\alpha=1,200(=2N^2)$. Only the first $N^2-1=99$ states
are occupied (the valence band). There are two discrete empty levels
deep into the Mott-Hubbard gap, one of which ($\alpha=100$) comes from
the valence band of the undoped AFM parent. Merons must be created in
vortex-antivortex pairs (for topological reasons). Each pair removes
two levels from the undoped AFM valence band. Thus, the valence band
remains spin paired, and the total spin of this excitation is
zero. This meron is a spinless, charged, bosonic collective excitation
of the doped antiferromagnet.  }
\end{figure}

\begin{figure}
\centering
\parbox{0.45\textwidth}{\psfig{figure=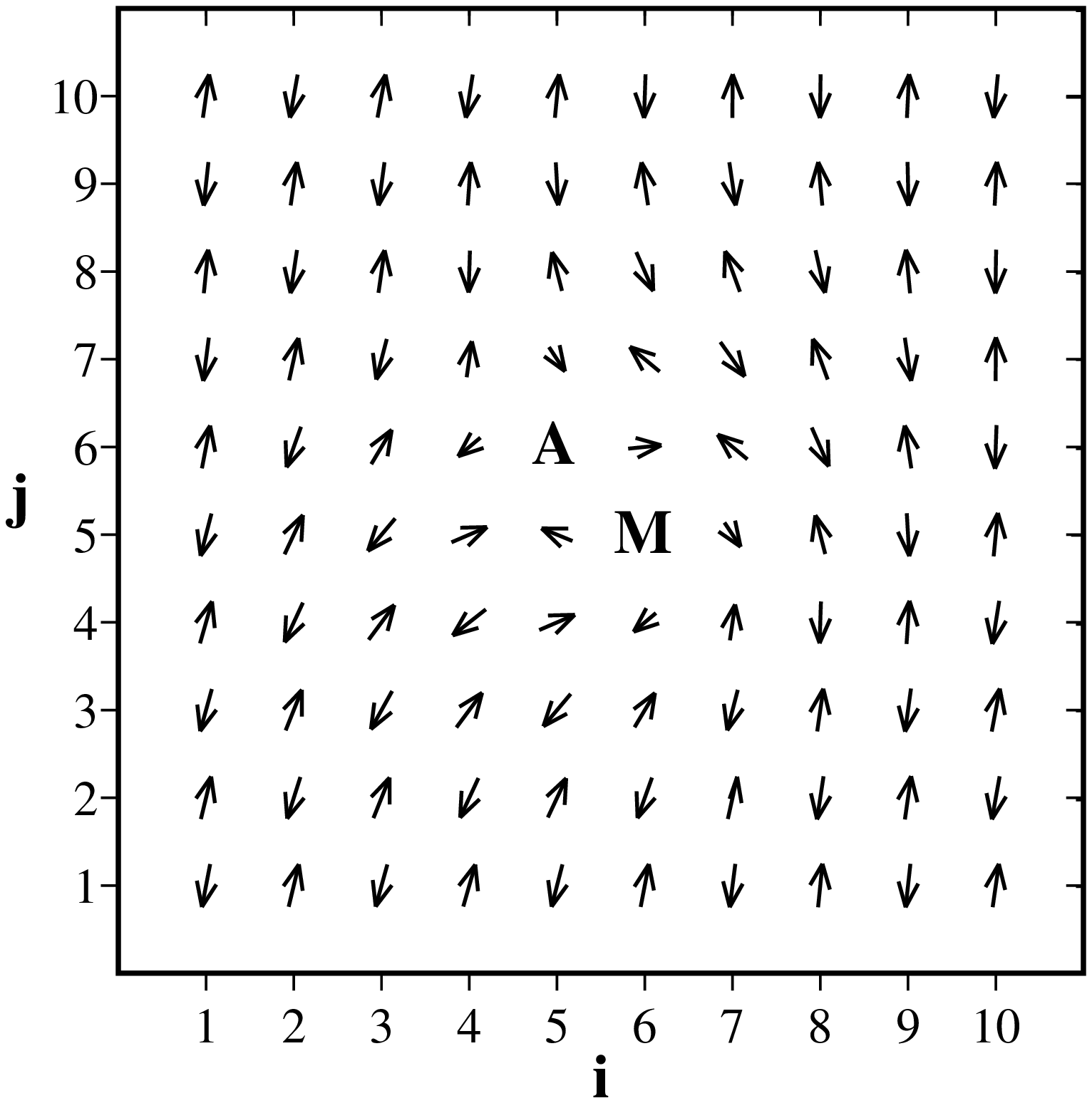,width=85mm}}
\parbox{0.45\textwidth}{\psfig{figure=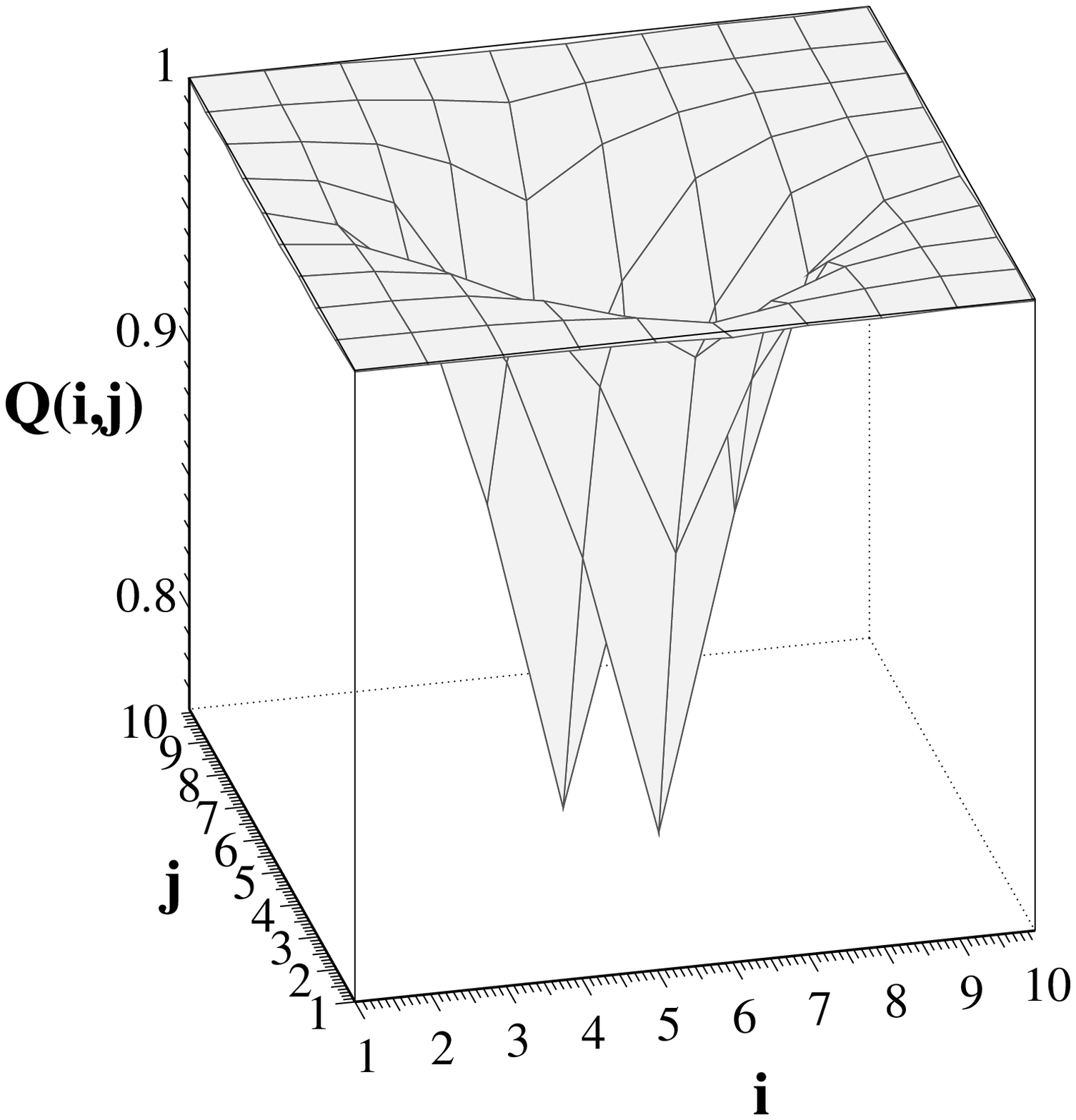,width=85mm}}
\caption{\label{fig2.25} (a). Self-consistent spin distribution for a
tightly bound meron-antimeron pair. The meron (M) and the antimeron
(A) are localized on neighboring sites. The total winding number of
the pair is zero. The magnetic AFM order is disturbed only on the
region where the vortices are localized.  The attraction between holes
is of topological nature and on long length scale is stronger than
unscreened Coulomb repulsion between charges.  (b). Self-consistent
charge distribution for a tightly bound meron-antimeron pair. The
doping charge is mostly localized on the two plaquettes containing the
meron and antimeron cores. The two holes localized in the vortex cores
are responsible for the fact that the meron-antimeron pair does not
collapse (due to Fermi statistics, it is impossible to have two holes
at the same site).  }
\end{figure}

\begin{figure}
\centering
\parbox{0.45\textwidth}{\psfig{figure=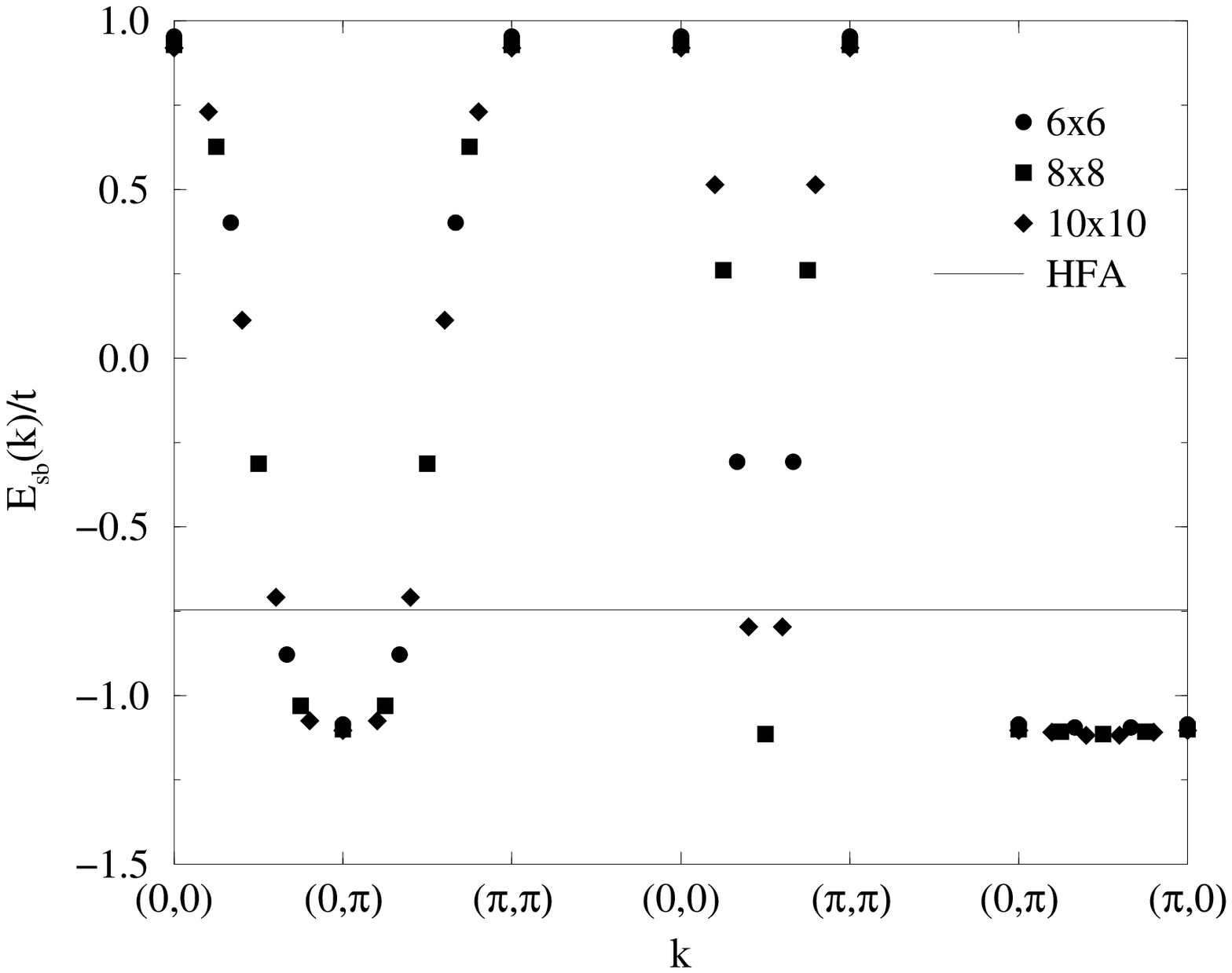,width=85mm}}
\parbox{0.45\textwidth}{\psfig{figure=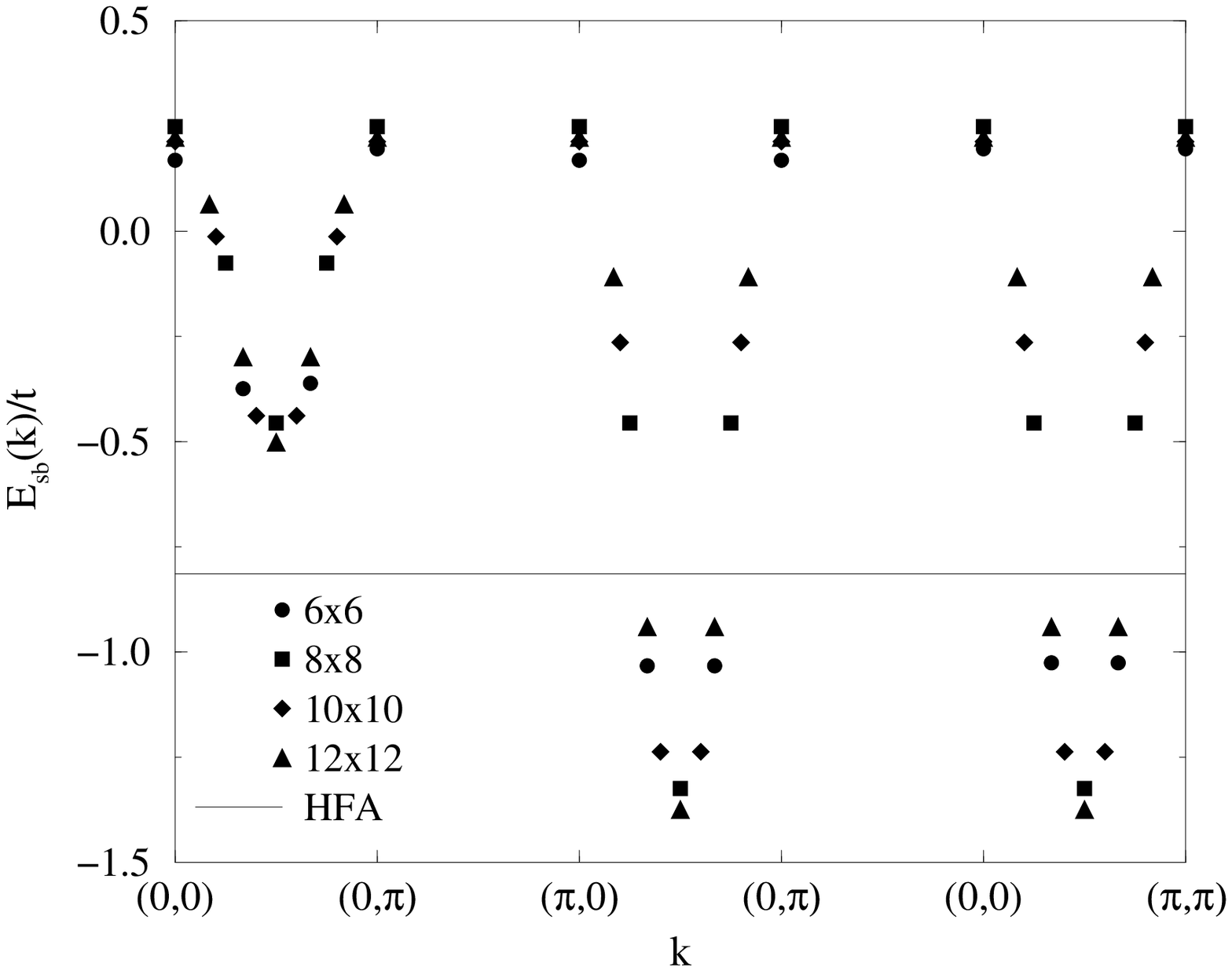,width=85mm}}
\caption{\label{fig3.20} Dispersion relation of the charged spin-bag
$E_{sb}(\vec{k})$ (in units of $t$) plotted along lines of high
symmetry in the Brillouin zone. The upper plot shows the dispersion
band of the spin-bag in the conventional model, while the lower one
shows the dispersion band of the spin-bag in the spin-flux
model. $U/t=5$ in both cases. Circles, squares, diamonds and triangles
show the results obtained from CI analysis of 6x6, 8x8, 10x10 and
12x12 lattices. We conclude that the results are already almost
converged, even for such small lattices. The full lines show the
excitation energy of the spin-bag at the static HF level.  }
\end{figure}

\begin{figure}
\centering
\parbox{0.45\textwidth}{\psfig{figure=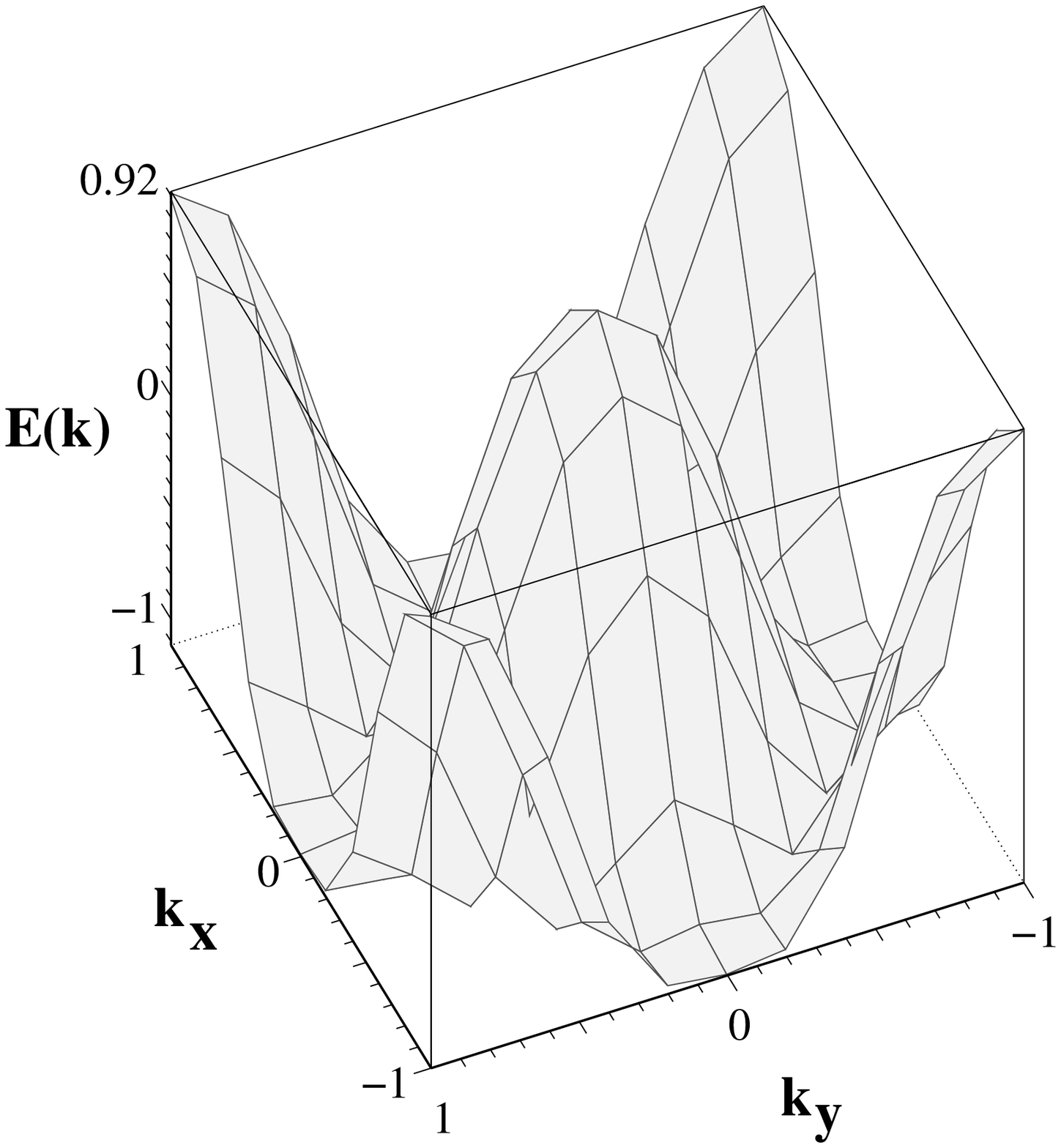,width=85mm}}
\parbox{0.45\textwidth}{\psfig{figure=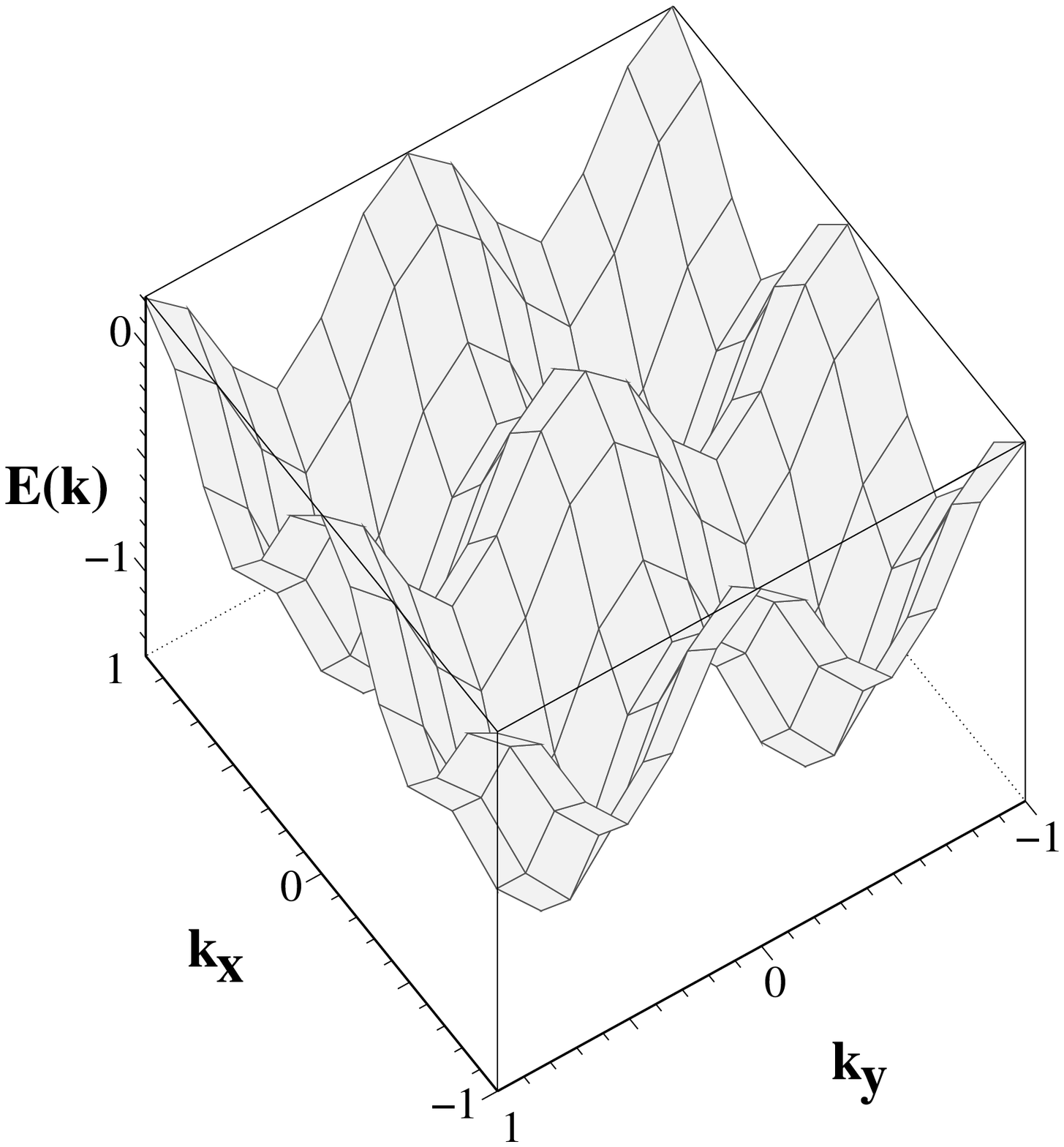,width=85mm}}
\caption{\label{fig3.21} Dispersion band of the spin-bag in the
conventional model (upper panel) and the spin-flux model (lower panel), for
$U/t=5$. We show the full 2D Brillouin zone ($k$ is measured in units
of $\pi/a$).  The spin-bag dispersion relations have the symmetry as
the dispersion relations of the underlying undoped AFM background,
shown in Fig. \ref{fig2.13}. While the $(\pi/2a, \pi/2a)$ point
corresponds to the minimum excitation energy of the spin-flux
spin-bag, in the conventional model all points along the $(0,\pi)$ to
$(\pi,0)$ have almost the same energy.}
\end{figure}

\begin{figure}
\centering
\parbox{0.45\textwidth}{\psfig{figure=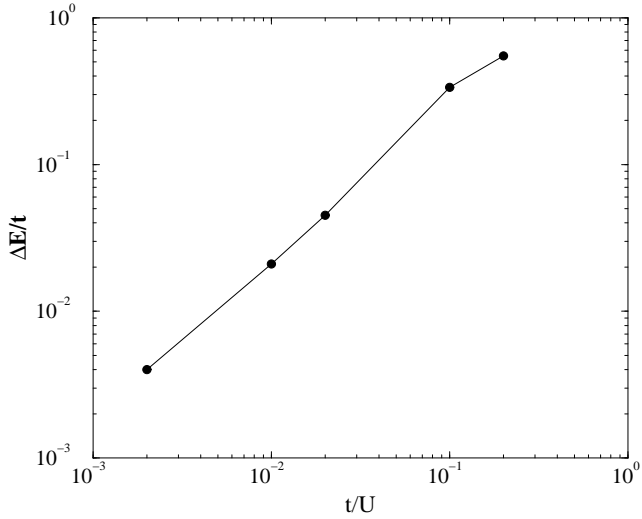,width=85mm}}
\caption{\label{fig3.24} The extra kinetic energy (in units of $t$)
gained by the spin-bag $\Delta E= E_{sb}({ \pi \over 2} ,{ \pi \over
2})- E_{sb}^{HF}$ (circles) as a function of $t/U$.  The log-log graph
clearly shows the linear dependence.  This is expected, since the
spin-bag must tunnel two sites to the next allowed position. This is a
second order hopping process and therefore this charged excitation is
rather immobile.  }
\end{figure}

\begin{figure}
\centering
\parbox{0.45\textwidth}{\psfig{figure=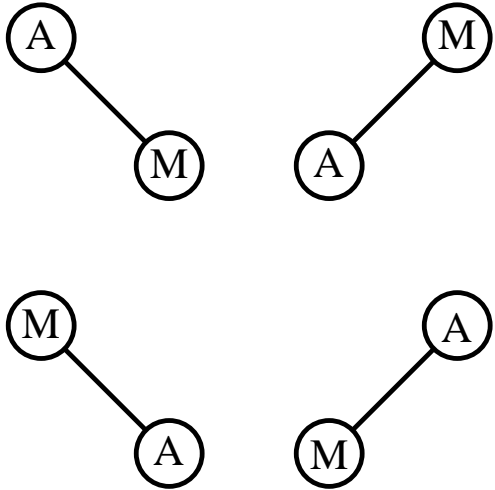,width=65mm}}
\caption{\label{fig3.28} Schematic representation of the four
different meron-antimeron pair configurations obtained through $\pi/2$
rotation about their fixed center of mass. The upper-left picture is a
schematic representation of the self-consistent meron-antimeron pair
shown in Fig.\ref{fig2.25}. }
\end{figure}

\begin{figure}
\centering
\parbox{0.45\textwidth}{\psfig{figure=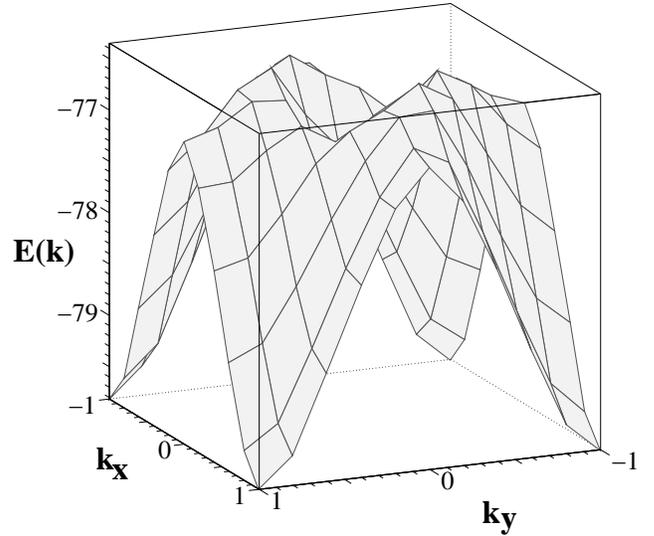,width=85mm}}
\caption{\label{fig3.25} Total energy $E(\vec{k})$ (in units of $t$)
vs. $\vec {k}$ of the lattice with an (un-rotated) meron-antimeron
pair as a function of the total momentum $\vec{k}$ of the pair.  The
momentum units are $\pi/a$.  The HF energy of the static
meron-antimeron pair is $-78.52t$.  Quantum hopping lowers the overall
energy of the pair by $1.29t$. Since the meron-antimeron configuration
is not rotationally invariant, the dispersion relation is also not
invariant to $\pi/2 $ rotations.  }
\end{figure}

\begin{figure}
\centering
\parbox{0.45\textwidth}{\psfig{figure=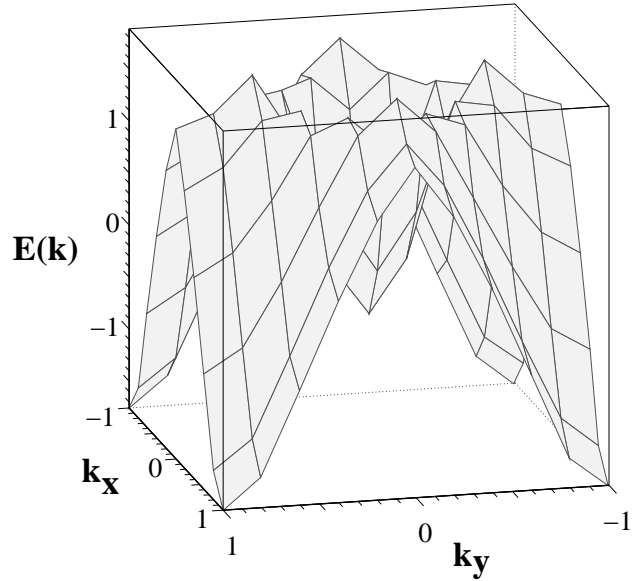,width=85mm}}
\caption{\label{fig3.26} The lowest energy dispersion band
$E_{pair}(\vec{k})$ (in units of $t$) as a function of the total
momentum $\vec {k}$ of the meron-antimeron pair. The momentum units
are $\pi/a$.  For convenience, the reference energy is taken to be the
static HF energy of the self-consistent meron-antimeron pair.  Quantum
hopping and rotation lowers the overall energy of the pair by
$1.75t$. The rotational symmetry of the dispersion relation is
restored (compared to Fig. \ref{fig3.25}). }
\end{figure}

\begin{figure}
\centering
\parbox{0.45\textwidth}{\psfig{figure=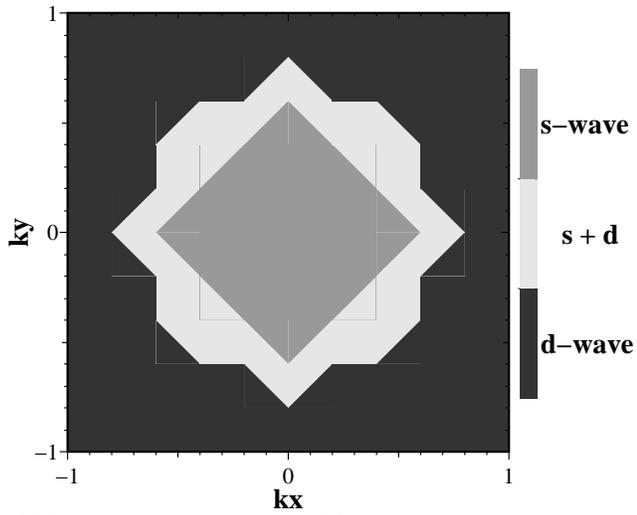,width=85mm}}
\caption{\label{fig3.27} The rotational symmetry of the
meron-antimeron wave-function as a function of the total momentum
carried by the pair (measured in units of $\pi/a$).  The outside
region (containing the absolute minima points $(\pi,\pi)$) has d-wave
symmetry ($J=2$), while the core region about the $(0,0)$ point has
s-wave symmetry ($J=0$). The intermediary area is a mix of s+d wave
symmetry.}
\end{figure}

\end{document}